\documentclass[traditabstract]{aa} % for the abstract without structuration 
\usepackage{graphicx}
\usepackage{txfonts}
\usepackage{wasysym}

\newcommand{\manuscript}{{\em Manuscript}}

\newcommand{\basic}{{\em WilhelmIV}}
\newcommand{\dists}{{\em WilhelmIV-Dist}}
\newcommand{\alts}{{\em WilhelmIV-Alt}}

\newcommand{\gtap}{\mathrel{\hbox{\rlap{\lower.55ex \hbox {$\sim$}}
                   \kern-.3em \raise.4ex \hbox{$>$}}}}
\newcommand{\ltap}{\mathrel{\hbox{\rlap{\lower.55ex \hbox {$\sim$}}
                   \kern-.3em \raise.4ex \hbox{$<$}}}}

\begin{document}
  \title{The star catalogue of Wilhelm IV, Landgraf von Hessen-Kassel
    \thanks{The full Tables \basic, \dists, and \alts\ (see Tables \ref{t:basic},
      \ref{t:dists} and \ref{t:alts}) are
      available in electronic form only at the CDS via anonymous ftp
      to cdsarc.u-strasbg.fr (130.79.128.5) or via
      http://cdsweb.u-strasbg.fr/cgi-bin/qcat?J/A+A/}}
  \subtitle{Accuracy of the catalogue and of the measurements}

  \author{Frank Verbunt\inst{1,2} and Andreas Schrimpf\inst{3} }

  \institute{
    Department of Astrophysics / IMAPP, Radboud University,
    PO Box 9010, 6500 GL Nijmegen, The Netherlands; \email{f.verbunt@astro.ru.nl}
	\and
    SRON Netherlands Institute for Space Research Utrecht, The Netherlands
        \and
	History of Astronomy and Observational Astronomy, Physics
    Department, Philipps Universit\"at, Renthof 5, 35032 Marburg,
    Germany; \email{andreas.schrimpf@physik.uni-marburg.de}
}

%  \date{Received \today / Accepted ??}
  \date{Draft hessenv9 \today}

  \abstract{We analyse a manuscript star catalogue by Wilhem\,IV, Landgraf
von Hessen-Kassel from 1586. From measurements of altitudes and of angles
between stars, given in the catalogue, we find that the measurement accuracy averages $26''$ for eight 
fundamental stars, compared to $49''$ of the measurements by Brahe.
The computation in converting altitudes to declinations and angles between
stars to celestial position is very accurate, with errors negligible with respect to the
measurement errors. Due to an offset in the position of the vernal equinox
the positional error of the catalogue is slightly worse than that of Brahe's
catalogue, but when correction is made for the offset -- which was known
to 17th century astronomers -- the catalogue is more accurate than that
of Brahe by a factor two. We provide machine-readable Tables of the catalogue.}

  \maketitle

\section{Introduction}

It is well known that the measurements by Tycho Brahe of the positions
of stars and planets in the sky were an order of magnitude more
accurate than those of earlier astronomers such as Ptolemaios (or
Hipparchos) and Ulugh Beg. As a result, Brahe's catalogue of positions
of the fixed stars became the new standard, with editions in 1598
(manuscript), 1602 (in Astronomiae Instauratae Progymnasmata) and in
1627  (by Kepler, Tabulae Rudolphinae). 
Machine-readable versions of these catalogues and an
analysis of their accuracy are provided by
Verbunt \&\ van Gent (2010, 2012). \nocite{verbunt10} \nocite{verbunt12}

It is less well known that Brahe was preceded by Wilhelm\,IV,  Landgraf von Hessen-Kassel,
whose measurements were equally accurate. 
As an aspiring astronomer, Brahe
visited the already experienced observer Wilhelm\,IV in April 1575,
and studied his instruments. In the correspondence that followed,
Wilhelm\,IV in turn profited from suggestions for improved measurement
accuracy by Brahe (Hamel 2002).\nocite{hamel02}

The star catalogue of Wilhelm\,IV was based mostly on work by
Wilhelm\,IV himself and by two people he employed: the 
mathematician-astronomer Christoph Rothmann,
and the brilliant clock-maker, instrument maker and mathematician
Jost B\"urgi. All three of them participated in the observations, with instruments 
made by B\"urgi. The computational work was presumably done by Rothmann and B\"urgi.
We describe the work at the observatory in Kassel and the historical context 
in a separate paper (Schrimpf \&\ Verbunt 2021).\nocite{schrimpf21} The
%compiled by the mathematician
%and astronomer Christoph Rot%hmann, who was employed at Kassel from
%December 1584 to 1590 (Gran%ada et al.\ 2003\nocite{granada03}). 
manuscript of the catalogue, almost ready for print, is available as a
scan via the on-line portal of the library of Kassel University
(Wilhelm IV 1587).\nocite{wilhelmiv_manuscript} Rothmann also
wrote a handbook for astronomers of his time describing
observational and computational methods he used for preparing the star
catalogue. The Latin manuscript for this handbook is available
via the on-line portal of the library of
Kassel University (Rothmann Handbook),\nocite{rothmann_hb} 
and has been printed by Granada et
al.\ (2003)\nocite{granada03}, with an introduction in German.

The catalogue only appeared in print, edited by Curtz \nocite{curtz66}
in 1666, long after the dissemination of Brahe's catalogue, and as a
result had a rather limited impact, even though the catalogue was
reprinted (with adaptations) by Hevelius (1690) \nocite{hevelius90}
and by Flamsteed (1725)\nocite{flamsteed25}. Hamel (2002), 
borrowing extensively from Wolf  (1878) and Repsold (1919),
\nocite{hamel02} has written a monograph on the astronomical research
in Kassel under Wilhelm\,IV, in which he mentions three earlier star
catalogues, with 58, 58 and 121 stars respectively. We intend to
compare these different catalogues, which mark the progress to the
higher accuracy, in a future paper. In the current paper we study the
manuscript version of the final catalogue, from 1587, which contains
387 entries.

Rothenberg (in Hamel 2002) compared this final catalogue with the SAO
star catalogue (SAO 1966), \nocite{sao} identified 361 stars securely,
found a possible counterpart for 7 and was unable to identify 14
stars. (Three repeated stars and two clusters of stars --
  see Table\,\ref{t:double} -- are not included in these numbers.)
Assuming an offset in right ascensions of 6$'$, he found systematic +
random errors 0\farcm22 + 1\farcm2 in right ascension and 0\farcm82 +
1\farcm5 in declination, respectively.  No details are provided about
individual (non-)identifications.  We refer to the final catalogue
below as \manuscript, identify all its entries by comparison with the
HIPPARCOS Star Catalogue (ESA 1997)\nocite{esa}, analyze the results
in detail and provide a machine-readable version of catalogue and
analysis, in three data files, \basic\ (see Table\,\ref{t:basic}),
\dists\ (Table\,\ref{t:dists}), and \alts\ (Table\,\ref{t:alts}).

\section{Description of \manuscript\ and its epoch\label{s:desc}}

The manuscript is organized by constellation, and lists all 1028
entries from the star catalogue by Ptolemaios, plus four added stars,
of which two in the Pleiades.
For all 1032 entries column one of the manuscript gives the Latin name
or description of the star, column seven the position in the catalogue
of Ptolemaios, or for the added stars the statement {\em in tabulis non
  extat} (not present in the table). Column eight gives the
magnitude. 21$\frac{1}{4}$ degrees have been added to the longitudes
from the catalogue of Ptolemaios to bring them to the equinox of the
new observations.

\begin{table}
\caption{Constellations in \manuscript\ and numbers of stars in them\label{t:consts}}
\centerline{\begin{tabular}{rcrr||rcrr}
\hline
\hline
C  & & $N$ & W  &C  && $N$ &  W  \\
\hline
 1 & UMi &  7  &  1 & 24 & Tau &  16 & 218   \\
 2 & UMa & 18 & 8 & 25 & Gem & 13 & 234 \\
 3 & Dra &  18  &26 & 26 & Cnc &  7 & 247\\
 4 & Cep & 10  &  44 &  27 & Leo &  9 & 254\\
 5 & Boo &  13 &  54  & 28 & Vir &  12 & 263\\
 6 & CrB &  4 &  67  & 29 & Lib &  2  & 275\\
 7 & Her &  10 &  71  &  30 & Sco &  6  & 277\\
 8 & Lyr &  8 &  81  & 31 & Sgr &  7  & 283\\
 9 & Cyg & 12  &  89 &  32 & Cap &  4 & 290\\
10 & Cas & 11  & 101 &  33 & Aqr &  9 & 294\\
11 & Per & 20 & 112 &34 & Psc & 16 & 303\\
12 & Aur &  8  & 132 &35 & Cet &  10 & 319\\
13 & Oph &  10 & 140 & 36 & Ori &  17 & 329\\
14 & Ser &  7  & 150 & 37 & Eri &  1  & 346\\
15 & Sge &  5  & 157 & 38 & Lep &  4  & 347\\
16 & Aql &  4  & 162 & 39 & CMi &  2  & 351\\
17 & Atn &  4  & 166  & 40 & CMa &  7 & 353\\
18 & Del &  5  & 170 & 41 & Nav &  3  & 360\\
19 & Equ &  3  & 175 & 42 & Hya & 17 & 363\\
20 & Peg &  12 & 178 &  43 & Crt &  3  & 380\\
21 & Tri &  3  & 190 & 44 & Crv &  4  & 383\\
22 & And & 16 & 193 & 45 & PsA & 1 & 387 \\
23 & Ari &  9  & 209  & \multicolumn{2}{c}{total} & 387 \\
\hline
\end{tabular}}
\tablefoot{For each constellation the columns give its sequence number $C$,
 the abbreviation we use, the number $N$ of stars in the constellation,
 and the sequence number $W$ of the first star in the constellation.
  Atn indicates Antinous, and Nav indicates Navis Argo.}
\end{table}

\begin{table}
\caption{Non-stellar, added, and repeated entries in
  \manuscript\label{t:double}}
\center{
\begin{tabular}{cc||cc}
\hline\hline
entry  & Cluster &  entry & Added star \\ \hline
 W\,112 (Per\,1) & h\&$\chi$\,Per  & W\,39 (Dra\,14) & RR\,UMi \\
 W\,247 (Cnc\,1) &Praesepe  & W\,232 (Tau\,15) & Alcyone \\
& &  W\,233 (Tau\,16) & Atlas \\
& &  W\,999 (Aql --) & $\epsilon$\,Aql \\
\hline\hline
\end{tabular}
\begin{tabular}{c c c}
 first entry & repeated entry & Star \\ \hline
W\,60 (Boo\,7)  & W\,80 (Her\,10) &$\nu^2$\,Boo  \\
W\,139 (Aur\,8) & W\,227 (Tau\,10) &$\beta$\,Tau \\
W\,302 (Aqr\,9) & W\,387 (PsA\,1) & $\alpha$\,PsA \\
\hline\hline
\end{tabular}
}
\end{table}

For a limited number of stars the intermediate columns provide new
measurements (angular distances between stars in columns two and three, and
meridional altitudes in column four) and the equatorial and ecliptic
coordinates derived from them (in columns five and six, respectively).

387 of the 1032 entries have newly determined ecliptic coordinates.
We collect these entries in \basic, and
number the stars in order of occurrence, with a
W number (see Table\,\ref{t:consts}).
The three stars that are (knowingly) repeated in Ptolemaios, also
occur twice in \manuscript, so that
\basic\ contains 384 independent entries.
Three of the four added stars have ecliptic coordinates.
We list these stars, together with the repeated entries and the entries with
non-stellar identifications  in Table\,\ref{t:double}.
For the fourth added star, in Aquila, two angular distances to other
stars, but no equatorial or ecliptic position, are given; we do not
include this star in \basic, and refer to it as W\,999. From the angular
distances given, we derive that it corresponds to $\epsilon$\,Aql.

For 384 entries (381 independent ones), the manuscript gives the
angular distance between the entry and one or more stars named in
column two. For all these entries the ecliptic coordinates are given
(in column 6).  We indicate the angular distance with $\phi$.  No
$\phi$ is given for W\,87, W\,88, and W\,224, but column six does give
their ecliptic position. For five stars the distance to a
  reference star is repeated, with the r\^oles of star and
  reference star reversed; we refer to the second occurrence as {\em
  reverse repeated angle}.  As we will see below, the position of
W\,224 (Aldebaran) is the anchor point from which absolute positions
for other entries are determined.

For 346 entries (343 independent ones), the manuscript
gives the altitude in the meridian passage and the
equatorial coordinates.  For stars that are always above the horizon
both superior and inferior altitudes are listed, indicated {\em
  superne} or {\em inferne}.  Table \ref{t:combinations}
details the frequencies of angular distances and altitudes in
\manuscript.

\begin{table}[b]
\caption{Frequency of different combinations of information given in
\manuscript, after removal of three repeated entries and
five reverse repeated angles \label{t:combinations}}
\begin{center}
\begin{tabular}{c|c|c|c|c||c|c}\hline\hline
 & 0\,$\phi$ & 1\,$\phi$ & 2\,$\phi$ & 3\,$\phi$ & \# &   \#\,$\phi$  \\
\hline
0\,h & 2 & 2 & 37 & 0 & 41 & 76 \\
1\,h & 1 & 257 & 53 & 0 & 311 & 363 \\
2\,h & 0 & 16 & 15 & 1 & 32 & 49 \\
\hline
\#  & 3 & 275 & 105 & 1 & 384 & \\
\hline
\hline
\#\,$h$& 1 & 289 & 83 & 2& &375$\backslash$488 \\
\hline
\end{tabular}
\end{center}
\tablefoot{Columns 2-5 sort the entries with respect to the number of
angular distances, rows 2-4 sort the entries with respect to the
number of altitudes (or equivalently, equatorial coordinates). 
Thus column 3, row 3 indicates that there are
257 entries with one angular distance and one altitude. 
Columns 6 and 7 give the total number of entries and angular
distances, respectively, from columns 2-5; rows 4 and 5 the
total number of entries and altitudes from rows 2-4. W\,999,
for which only two angular distances are given, is not included in
the numbers.
}
\end{table}

All angles in \manuscript\ are given in degrees and sexagesimal fractions; 
for latitudes or declinations S or M is added for Septentrionalis
(North) or Meridionalis (South), respectively. For the
machine-readable tables we convert the fraction into seconds:
\begin{equation}
q = G_q + \frac{M_q+F_q}{60};\quad S_q=60F_q\,;\quad
 q = \phi,h,\alpha,\delta,\beta
\label{e:gms}\end{equation}
where $\phi,h,\alpha,\delta,\beta$ refer to angular distance, altitude
in the meridian, right ascension,  declination and ecliptic latitude,
respectively. The values of $F_q$ in the catalogue are such that $S_q$ is always integer,
one of the 14 following:  0,6,10,12,15,20,24,30,36,40,45,48,50,54. Remarkably,
18 (=3/10) and 42 (=7/10) are not used.
For ecliptic longitude, a zodiacal sign is used, which we convert into
a number $Z$, from 1 for Aries \aries\ to 12 for Pisces \pisces:
\begin{equation}
\lambda = (Z-1)30+G_\lambda+\frac{M_\lambda+F_\lambda}{60};\quad S_\lambda=60F_\lambda
\label{e:longitude}\end{equation}

In the preamble of the 121-star catalogue,  prepared in 1586,  
the equatorial and ecliptic
coordinates of Aldebaran are given and said to form the fundament of
the coordinates of all other stars in the catalogue. The coordinates
of Procyon are also given and said to be very accurate. A 
translation of this Latin preamble is given by Schrimpf \&\ Verbunt 
(2021).\nocite{schrimpf21} The coordinates of Aldebaran and Procyon in
this preamble (Table\,\ref{t:aldeproc}) are identical to those given
in \manuscript, with the exception of the longitude of Procyon, for which
$F_\lambda$ is 15 larger in \manuscript.

For geophysical latitude $\phi_G$, the declination of a star follows
immediately from its altitude at the meridional culmination with
Eq.\,\ref{e:geophim}. Conversely, from the altitudes in \manuscript\ 
(54\degr17\arcmin\ for Aldebaran and 44\degr54\arcmin\ for Procyon) and
their declinations we can derive the value of $\phi_G$ used in
constructing the catalogue. For both Aldebaran and Procyon exactly the
same value $\phi_G=51\degr19\arcmin$ follows. This agrees with
the value given in Chapter\,10 of Rothmann's Handbook.

Ecliptic coordinates are computed from equatorial coordinates with
Eqs.\,\ref{e:eqtoecl},\ref{e:eqtoecb}, which require a value for the obliquity $\epsilon$.
From the equatorial and ecliptic coordinates of Aldebaran
and Procyon in the preamble and \manuscript\ we can compute the
obliquity used in constructing the catalogue. The ecliptic coordinates
derived from the equatorial coordinates with
$\epsilon=23\degr31\arcmin$ exactly match $\lambda$ and
$\beta$ of Aldebaran in the preamble and \manuscript, and
closely match the values for  Procyon, with $F_\lambda$ too large
by 19 in the preamble (invisible due to its rounding to minutes),
reduced to just 4 in \manuscript, and by only 2\arcsec\ for
$\beta$.
Our analysis in Sect.\,\ref{s:eqtoecl} of the full catalogue confirms
this value for $\epsilon$, somewhat surprisingly since
values of $\epsilon=23\degr31\arcmin24\arcsec$ and
 $\epsilon=23\degr31\arcmin30\arcsec$ are given in Chapters 10
and 12 of the Handbook, respectively.

\manuscript\ does not give the equinox. It is based on observations
made in 1586 and 1587, and presumably written in 1587,
which suggests an equinox of 1586 (Wolf 1878).\nocite{wolf78}
The manuscript of the catalogue for 121 stars
gives an equinox of 1586. The stellar positions of the stars
in this catalogue are, with few exceptions, identical to those of 
\manuscript. 
Therefore we use 1 January 1586 (old style) =  JD\,2300345,
as the epoch and the equinox of \manuscript.

The positions of the entries in \manuscript\ are anchored on the
position of Aldebaran. Brahe determined the right ascension of this 
star to be 6 arcmin smaller than the value used in Kassel, and 
discussed this discrepancy with  Rothmann and Wilhelm\,IV
(see Wolf 1878 \nocite{wolf78} and Hamel 2002
\nocite{hamel02} for more detail).
From the way the positions of the other stars are derived from
the measurements (see Eqs.\,\ref{e:geophis}-\ref{e:geophim} and
Eq.\,\ref{e:alpha}), this would imply that
all right ascensions in \manuscript\ are too large by 6 arcmin,
if Brahe's value is correct.

\begin{table}
\caption{Equatorial and ecliptic positions of Aldebaran and Procyon
\label{t:aldeproc}}
{\center
\begin{tabular}{l@{{\hspace*{0.05cm}}}r@{{\hspace*{0.12cm}}}rr@{{\hspace*{0.12cm}}}r@{{\hspace*{0.12cm}}}c||c@{{\hspace*{0.12cm}}}r@{{\hspace*{0.12cm}}}r@{{\hspace*{0.12cm}}}rr@{{\hspace*{0.12cm}}}r@{{\hspace*{0.12cm}}}r@{{\hspace*{0.12cm}}}c}
\hline\hline
star & $G_\alpha$ & $M_\alpha$ & $G_\delta$ & $M_\delta$ &H  & Z & $G_\lambda$ & $M_\lambda$ &
 $F_\lambda$ & $G_\delta$ & $M_\delta$ & $S_\delta$ & H\\
\hline
Aldebaran & 63 & 10 & 15 & 36 & S & \gemini & 4 & 06 & 00 & 5 & 31 &
45 & M \\
\multicolumn{6}{l||}{\phantom{MM}computed from equatorial}
  & \gemini & 4 & 06 & 00 & 5 & 31 & 45 & M\\
\phantom{MM}HIP & 63 & 4 & 15 & 36 & S & \gemini & 4 & 00 & 21 & 5 & 29 &
49 & M \\ 
Procyon    &109 & 30 &   6 & 13 & S & \cancer & 20 & 11 & 15 & 15 & 56 & 20
& M \\
\multicolumn{6}{l||}{\phantom{MM}computed from equatorial}
  & \cancer & 20 & 11 & 19 &15 & 56 & 22 & M \\
\phantom{MM}HIP & 109 & 24 & 6 & 13 & S & \cancer & 20 & 04 & 41 & 15 & 56
&06 & M\\
\hline
\end{tabular}
}
\tablefoot{For each star the first line has the coordinates from
  \manuscript, the second line the ecliptic coordinates computed
from the equatorial coordinates in \manuscript\ with
$\epsilon=23\degr31\arcmin$, and the third line the coordinates
in 1586  computed from HIPPARCOS  data.}
\end{table}

Comparison with the values computed from HIPPARCOS-2 data
(Van Leeuwen 2007) \nocite{leeuwen07} show that Brahe's criticism was 
justified (Table\,\ref{t:aldeproc}).

\section{Identification\label{s:identification}}

The method of identification of the stars largely follows the
procedure outlined in Verbunt \&\ van Gent (2010), with two
modifications.\nocite{verbunt10} For the computation of the star
positions at the epoch of the catalogue, we use data from HIPPARCOS-2,
the reanalysis of the HIPPARCOS catalogue by Van Leeuwen
(2007).\nocite{leeuwen07} Furthermore, we take into account the
limited spatial resolution of the naked human eye, by merging stars
within $2'$ of one another.  The visual magnitudes are from the
original HIPPARCOS catalogue (ESA 1997).\nocite{esa} On the basis of
this, we prepare a version of the HIPPARCOS catalogue converted to the
epoch 1 Jan 1586, as detailed in Appendix\,\ref{s:hipparcos}. When
referring to the HIPPARCOS Catalogue below, this converted version is
meant.

In view of the systematic offset of 6$'$ in right ascension noted by Brahe, confirmed in
Table\,\ref{t:aldeproc}, and  further discussed in Section\,\ref{s:accurcat},
we search for counterparts with
\begin{equation}
\Delta\alpha=\alpha-\alpha_\mathrm{HIP};\quad \Delta\alpha_\mathrm{c}=\Delta\alpha-6';
\quad \Delta\delta=\delta-\delta_\mathrm{HIP}
\label{e:eqerrors}\end{equation}
by looking for the entry with the smallest value of
\begin{equation}
\Delta_c
=2\arcsin\sqrt{\sin^2\frac{\Delta\delta}{2}+
\cos\delta\cos\delta_\mathrm{HIP}\sin^2\frac{\Delta\alpha_\mathrm{c}}{2}}
\label{e:eqdelta}\end{equation}
Eq.\,\ref{e:eqdelta}, the haversine function, is accurate also for very small angles.
For stars for which the catalogue gives ecliptic coordinates only, we convert the ecliptic 
coordinates to equatorial coordinates, using the value for the obliquity that
was used in Kassel, $\epsilon=23^\circ31'$. 
In virtually all cases the nearest star is the counterpart, in some cases a brighter
star at slightly larger distance is the counterpart.  For W\,112 and W\,247
the counterpart is a star cluster (see Table\,\ref{t:double}). 

\begin{table*}
\caption{First and eigth line of the machine-readable table \basic.\label{t:basic}}
%                      wc     con nG     M     SG     M     S     HZ     g      M    SG     M     S      H     VH     F      V
\center{
\begin{tabular}{c@{\hspace*{0.3cm}}c@{ }l@{ }r@{\hspace*{0.3cm}}c@{ }c@{ }c@{\hspace*{0.3cm}}c@{ }c@{ }c@{ }c@{\hspace*{0.3cm}}c@{ }c@{ }c@{ }c@{\hspace*{0.3cm}}c@{ }c@{ }c@{ }c@{ }c@{\hspace*{0.3cm}}r@{  }c@{ }c@{\hspace*{0.3cm}}r@{\hspace*{0.2cm}}r@{\hspace*{0.2cm}}r@{\hspace*{0.3cm}}r@{\hspace*{0.2cm}}r@{\hspace*{0.2cm}}r@{\hspace*{0.3cm}}l} \\ 
\hline\hline
 W & C & Con & i & $G_\alpha$& $M_\alpha$ & $S_\alpha$ & $G _\delta$&
 $M_\delta$ &$S_\delta$& $H_\delta$ & $Z$ & $G_\lambda$ & $M_\lambda$ & $S_\lambda$ &$G_\beta$ & $M_\beta$
 & $S_\beta$ & $H$ & $V$ & HIP\phantom{i} & I & $V_\mathrm{H}$ & $\Delta\alpha$ &
 $\Delta\delta$ & $\Delta_c$ & $\Delta\lambda$ & $\Delta\beta$ & $\Delta$ & description\\
\hline
  1 &  1 & UMi & 1 & 005 & 46 & 00 &  87 &04& 00& S &  3& 22& 48& 10&  66& 01& 15& S
&  2&  11767&  1&  2.0& 11.2 & 0.3   & 0.4 & 1.4  & $-$1.9 & 2.0  &Stella Polaris\\                    
\ldots \\
 8 & 2 & UMa & 1 &&&&&&&& 4 & 17 & 15 & 40 & 40 & 12 & 00 & S & 4 & 41704 & 1 & 3.3 
& 5.6 & 0.0 & 0.2 & 3.0 & $-$0.7 & 2.4 & In rostro. \\
          \\         
\end{tabular}}
\tablefoot{For explanation of the columns see Sect.\,\ref{s:machine}}
\end{table*}

\section{Machine-readable tables\label{s:machine}}

The machine-readable version of \manuscript\ is 
divided intro three tables, to minimize the number of 
empty slots. 
\basic\ contains the equatorial and ecliptic coordinates
resulting from the observations in Kassel 
and the magnitude, that is columns one, five, six and eight
of \manuscript\ (Table\,\ref{t:basic}).
In column 1 it gives the sequence number W of the entry,
in columns 2 and 3 the sequence number C and name of the 
constellation, in column 4 the sequence number $i$ of the star 
within the constellation, in columns 5 to 11 
the equatorial coordinates expressed in $G_\alpha$, $M_\alpha$, $S_\alpha$
and in $G_\delta$, $M_\delta$, $S_\delta$ and hemisphere indicator 
$H_\delta$ (Eq.\,\ref{e:gms}), in columns 12 to
19 the ecliptic coordinates in zodiacal sign $Z$ plus $G_\lambda$,
$M_\lambda$, $S_\lambda$ and $G_\beta$, $M_\beta$, $S_\beta$ and 
hemisphere indicator $H$ (Eqs.\,\ref{e:gms}, \ref{e:longitude}), 
and in column 20 the magnitude $V$. 

Columns 21 to 26 give the result of our analysis with,
  respectively, the HIPPARCOS identification, the flag I giving the
  quality of the identification, the magnitude $V_H$ of the HIPPARCOS
  star, the positional errors $\Delta\alpha$,
  $\Delta\delta$ and $\Delta_\mathrm{c}$ in arcminutes, according to
  Eqs.\,\ref{e:eqerrors} and \ref{e:eqdelta}.  We also list positional
  errors when no correction for the offset in right ascension is
  applied. For this we convert the equatorial coordinates
  $\alpha_\mathrm{HIP}$,\,$\delta_\mathrm{HIP}$ to ecliptic
  coordinates $\lambda_\mathrm{HIP}$,\,$\beta_\mathrm{HIP}$ with the
  value of the obliquity in 1586 according to modern theory
  $\epsilon(1586)=23\fdg4931$ (Seidelman 1992,
  Eq.\,3.222-1).\nocite{seidelman92} For each HIPPARCOS entry with
  $V\leq5.3$ we compute the angular distance $\Delta$ between the
  position in the old catalogue and the position computed from 
  HIPPARCOS data with
\begin{equation}
\Delta\lambda = \lambda-\lambda_\mathrm{HIP};\quad \Delta\beta=\beta-\beta_\mathrm{HIP}
\label{e:eclerrors}\end{equation}
and
\begin{equation}
\Delta
=2\arcsin\sqrt{\sin^2\frac{\Delta\beta}{2}+
\cos\beta\cos\beta_\mathrm{HIP}\sin^2\frac{\Delta\lambda}{2}}
\label{e:ecldelta}\end{equation}
Columns\,27 to 29 give $\Delta\lambda$, $\Delta\beta$ and $\Delta$
computed with these equations, in arcminutes.
Column 30  gives the name or description of the star as given
in column 1 of \manuscript. We consider all our identifications secure.
The identifications flags indicate identifications with 1 =  nearest star (with $V<5.3$),
2 = not nearest star, 3 = merged HIPPARCOS star (see Appendix A), 9 = star cluster,
6 = repeated entry.

\dists\ gives the angles measured between entries and reference
stars, i.e. columns one to three of \manuscript\
(Table\,\ref{t:dists}). All reference stars
listed in the second column also occur in the first column, and thus
can be identified with a W number. 
Column 1 of \dists\ gives the sequence number W for the entry,
column 2 idem $W_R$ for the reference star, column 3 the flag F
for repeated entries (1 for a repeated entry, 2 for a reversed repeated angle),
columns 4-6  the angular distance $\phi$ between entry and 
reference star in terms
of $G_\phi$, $M_\phi$ and $S_\phi$ (Eq.\,\ref{e:gms}), column 7
the difference (in arcminutes) between this angle and the value 
$\phi_\mathrm{HIP}$ computed from  HIPPARCOS-2 data, 
and column 8 the name or description of the reference star.
Columns 9 and 10 of the machine-readable file 
(not shown in Table\,\ref{t:dists}) give the
HIPPARCOS numbers of star and reference star.

\alts\ contains the altitude(s) for each star, i.e. column 4 of 
\manuscript. Column 1 of \alts\ gives the sequence number $W$,
column 2 the repeat flag (1 for a repeated entry), columns 3-6
the altitude $h$ in terms of $G_h$, $M_h$, $S_h$ and indicator 
S for superior, I for inferior or M for
southern (meridional) culmination. 
Column 7 gives the difference between the tabulated altitude $h$ and the
true altitude $h_\mathrm{HIP}$ computed from HIPPARCOS-2 data
with Eqs.\,\ref{e:geophis} -- \ref{e:geophim} with the modern
value $\phi_\mathrm{G}=51\fdg31367$. Column 8 gives the
HIPPARCOS number of the star.

\begin{table}
\caption{First lines of the machine-readable table \dists, with
  examples of a repeat entry (F=1) and of a reverse repeat entry (F=2). \label{t:dists}}
\center{
\begin{tabular}{rrcc@{ }c@{ }ccl}
\hline\hline
 W & W$_\mathrm{r}$ & F &\multicolumn{3}{c}{$\phi$}& 
$\phi-\phi_\mathrm{HIP}$ & reference star\\
    &  & & $G_\phi$ & $M_\phi$ & $S_\phi$ & (\arcmin) \\
\hline
  1&  92 & 0 & 44& 40 &12 & $-$1.1 & Cauda Cygni \\
  1& 104 & 0 & 28& 35& 45& $-$0.8 & Ad coxas Cassiopeae\\
  1& 132 & 0 & 43& 23& 30& $-$0.9 & Capella\\
  2& 132 & 0 & 47& 21& 00& $-$0.9 & Capella\\
\multicolumn{2}{c}{\ldots} \\
 60 & 25 & 0 & 20 & 04 & 00 &$-$0.8 & Ulti: Caudae Urs: ma:\\
\multicolumn{2}{c}{\ldots} \\
 80 & 25 & 1 & 20 & 04 & 00 &$-$0.8 & Ult: Caudae Urs: ma:\\
\multicolumn{2}{c}{\ldots} \\
121 & 132 & 0 & 23 & 39 & 20 & $-$1.2 & Capella \\
\multicolumn{2}{c}{\ldots} \\
132 & 121 & 2 & 23 & 39 & 20 & $-$1.2 & Cap: Medusae \\
\hline
\end{tabular}}
\tablefoot{For explanation of the columns see Sect.\,\ref{s:machine}.
Two angles are given in \manuscript\ for a star
which is not itself an entry; this star is listed in \dists\ with W = 999.
The machine-readable table also gives the HIPPARCOS numbers for
W and W$_\mathrm{r}$.}
%\end{table}
%\begin{table}
\caption{First lines of the machine-readable table \alts.\label{t:alts}}
\center{
\begin{tabular}{ccc@{ }c@{ }c@{ }ccr}
\hline\hline
 W &  F &\multicolumn{3}{c}{$h$}&  & $h-h_\mathrm{HIP}$ & HIP\phantom{7} \\
    &  & $G_h$ & $M_h$ & $S_h$ & & (\arcmin) \\
\hline
1 & 0 & 54 & 15 & 00 & S & $-$0.1 & 11767\\
1 & 0 & 48 & 23 & 00 & I & \phantom{$-$}0.5 & 11767 \\
4 & 0 & 65 & 29 & 00 & S & \phantom{$-$}0.8 & 72607\\
4 & 0 & 37 & 09 & 00 & I & $-$0.4 & 72607\\
\multicolumn{2}{c}{\ldots} \\
 80 & 1 & 81 & 01 & 00 & M & $-$1.1  & 76041\\
\hline
\end{tabular}}
\tablefoot{For explanation of the columns see Sect.\,\ref{s:machine}}
\end{table}

\section{Accuracy of the computations\label{s:computation}}

The accuracy of computation can be determined from tabulated
numbers in \manuscript\, that are derived from other tabulated numbers. 

\begin{figure}
\centerline{\includegraphics[angle=0,width=\columnwidth]{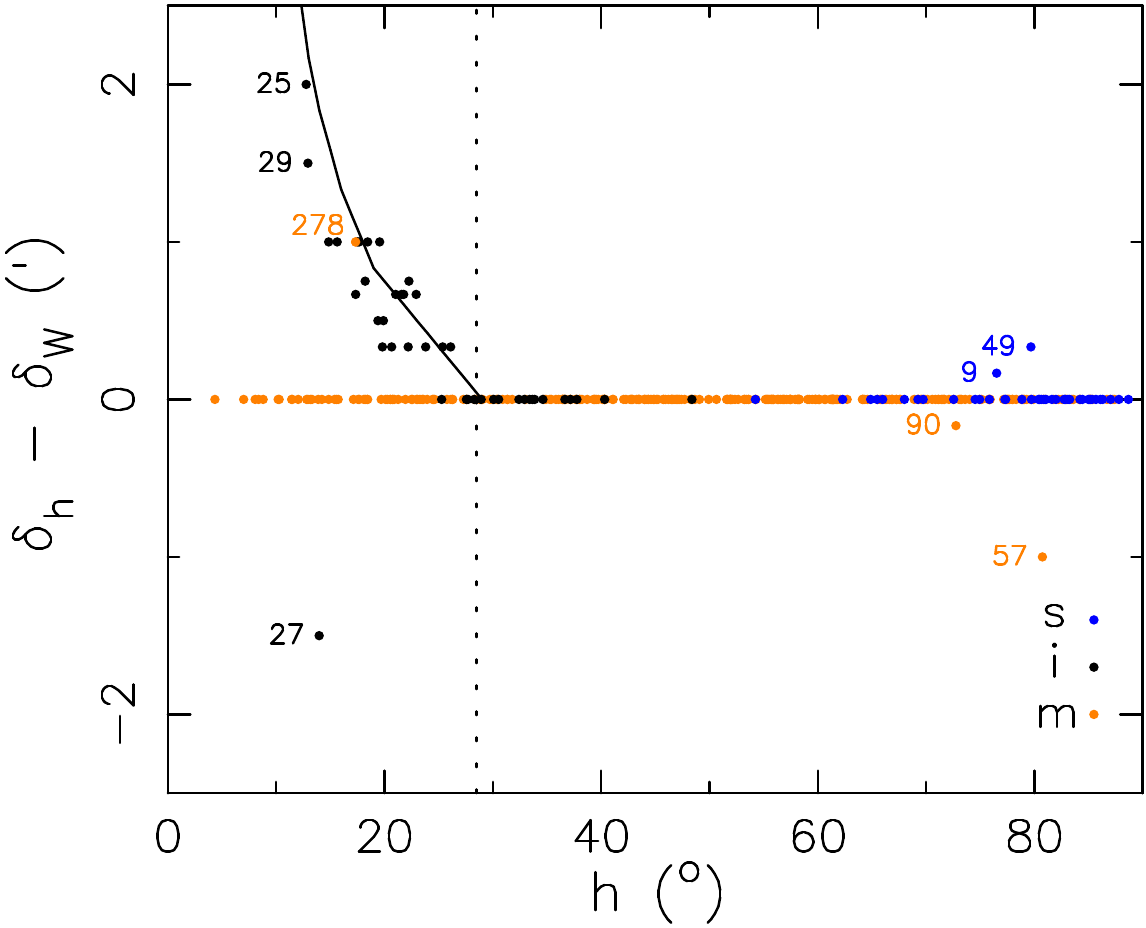}}

\caption{The difference between the declination $\delta_h$ computed
  from the altitude with Eqs.\,\ref{e:geophis}-\ref{e:geophim} and the
 declination $\delta_\mathrm{W}$ listed in \manuscript, for 
 $\phi_G=51\degr19\arcmin$.
 With few exceptions, the computations are exact for superior and
 southern culminations. The values $\delta_h$ from inferior culmination
 are systematically too high for low altitudes. Values for which 
$\delta_\mathrm{h}\neq\delta_\mathrm{W}$ are labelled with W for 
those entries for which only one altitude is listed, and for W\,9.
The solid curve gives the refraction $R_\mathrm{R}$ according to Rothmann.
Stars to the right of the vertical dashed line have $R=0$ according to Rothmann.
\label{f:phigeo}}
\end{figure}

\begin{table}
\caption{Refraction as a function of altitude according to Rothmann\label{t:refract}}
\center{
\begin{tabular}{cr|cr|cr|cr}
\hline\hline
\hline
$h$ & $R_\mathrm{R}$\phantom{41} & $h$ & $R_\mathrm{R}$\phantom{41} &
$h$ & $R_\mathrm{R}$\phantom{41} & $h$ & $R_\mathrm{R}$\phantom{1} \\
\hline
 2\degr & 13\arcmin40\arcsec &  9\degr &  4\arcmin40\arcsec & 16\degr &  1\arcmin20\arcsec & 23\degr &  30\arcsec \\
 3\degr & 12\arcmin20\arcsec & 10\degr &  3\arcmin50\arcsec & 17\degr &  1\arcmin10\arcsec & 24\degr &  25\arcsec \\ 
 4\degr & 11\arcmin00\arcsec & 11\degr &  3\arcmin10\arcsec & 18\degr &  1\arcmin00\arcsec & 25\degr &  20\arcsec \\
 5\degr &  9\arcmin35\arcsec & 12\degr &  2\arcmin40\arcsec & 19\degr & 50\arcsec & 26\degr &  15\arcsec \\ 
 6\degr &  8\arcmin10\arcsec & 13\degr &  2\arcmin10\arcsec & 20\degr & 45\arcsec & 27\degr &  10\arcsec \\
 7\degr &  6\arcmin50\arcsec & 14\degr &  1\arcmin50\arcsec & 21\degr & 40\arcsec & 28\degr &  5\arcsec \\
 8\degr &  5\arcmin40\arcsec & 15\degr &  1\arcmin35\arcsec & 22\degr
 & 35\arcsec & $\geq$29\degr & 0\phantom{\arcsec}\\
\hline
\end{tabular}}
\end{table}

\subsection{Altitudes and declinations\label{s:altidecs}}

The relations between the geographical latitude $\phi_\mathrm{G}$ of
the observer, the declination $\delta$ of a star, and its altitude $h$
at meridian passage for culminations above (superior) or below
(inferior) to the celestial north pole in the northern direction, and
culminations in the southern (meridional) direction are, respectively:
\begin{eqnarray}
\delta &=& \phi_\mathrm{G} -h +\frac{\pi}{2}
 \hspace*{2cm}\mathrm{superior\ culmination} \label{e:geophis}\\
\delta &=& -\phi_\mathrm{G} +h +\frac{\pi}{2} 
\hspace*{1.80cm}\mathrm{inferior\ culmination}\label{e:geophii}\\
\delta &=&\phi_\mathrm{G}  +h -\frac{\pi}{2}
\hspace*{1.99cm}\mathrm{meridional\ culmination} \label{e:geophim}
\end{eqnarray}
In these equations $h$ is the true altitude.
The observed or apparent altitude $h_\mathrm{a}$ is affected by
atmospheric refraction $R$:
\begin{equation}
h_\mathrm{a} = h + R 
\label{e:altsapp}\end{equation}
The values for refraction used by Rothman, $R_\mathrm{R}$ are listed
in Table\,\ref{t:refract}; according to Rothmann no refraction occurs for
$h\geq29\degr$.
As noted above, all entries in \manuscript\ for which an altitude is
listed have equatorial coordinates. A single altitude is listed
for 311 entries, viz.\ 293, 10 and 8 entries from meridional, superior  and inferior culmination,
respectively, for 32 entries the altitudes are listed both for superior and inferior
culmination.
If we enter these altitudes and
$\phi_\mathrm{G}=51\degr19\arcmin$ into Eqs.\,\ref{e:geophis}-\ref{e:geophim}
we obtain declinations $\delta_\mathrm{h}$ equal to the catalogued
declinations $\delta_\mathrm{W}$  listed in column\,5 of \manuscript, for all
but  3 and 2 values obtained from meridional and superior culmination,
respectively, as shown in Fig.\,\ref{f:phigeo}. For meridional altitudes $h<29\degr$
this implies that the tabulated altitudes are true altitudes, corrected
for refraction -- or at least were considered as such in the
computation of $\delta$.
24 of the inferior altitudes  lead to a different declination
than given in the catalogue, i.e.\ $\delta_\mathrm{h}\neq\delta_\mathrm{W}$.
The differences of declinations derived from altitudes of inferior  culmination
with the catalogued values roughly follow the atmospheric refraction,
leading to the assumption that these are apparent altitudes.
However, the correspondence is not exact, which
we are inclined to ascribe to rounding and / or to copying errors.

Thus, the 378 altitudes listed in \manuscript, and collected by us in
\alts, lead to 343 independent entries with declinations: the other altitudes
are 3 repeated entries and 32  altitudes of inferior culminations which
are not used.

\subsection{Angles between stars and right ascensions}

The standard equation for the angular distance $\phi$
between two stars may be rewritten in equatorial coordinates as
\begin{equation}
\alpha=\alpha_\mathrm{r} + \arccos\left(
\frac{\cos\phi-\sin\delta\sin\delta_\mathrm{r}}{\cos\delta\cos\delta_\mathrm{r}}\right)
\label{e:alpha}\end{equation} 
showing that the right ascension $\alpha$  of a star can be computed from
its declination $\delta$ if the angle $\phi$ is known to a reference star with known
equatorial coordinates $\alpha_\mathrm{r}$, $\delta_\mathrm{r}$. 
\manuscript\ contains 500 values for $\phi$ between entries, of which
410 are independent and between stars with listed equatorial coordinates.
Of these 410 entries, 273 have $\phi$ with one reference star,
67 with two reference stars (i.e.\ 134 values of $\phi$), and 1 (W\,1) with three reference stars.

To retrace the steps presumably taken in Kassel, we use the
tabulated values of $\phi$ in subsequent iterations.  In the first
iteration, we have one reference star, Aldebaran, for which angles are
given to 21 entries with declination known from an altitude
measurement. The first iteration thus allows the computation of the
right ascension for these 21 entries. 7 of these, that occur in the
second column of \dists, can be used as a reference star in the second
iteration. After eight iterations, a total of 64 stars have been used
as reference star, and the equatorial coordinates of 342 stars are
known. Only one star with equatorial coordinates in \manuscript,
W\,288, is not found in these iterations; the angle $\phi$ in \dists\
between this star and W\,291 does not help as no altitude $h$ and thus
no equatorial coordinates are listed for W\,291.

\begin{figure}
\centerline{\includegraphics[angle=0,width=0.8\columnwidth]{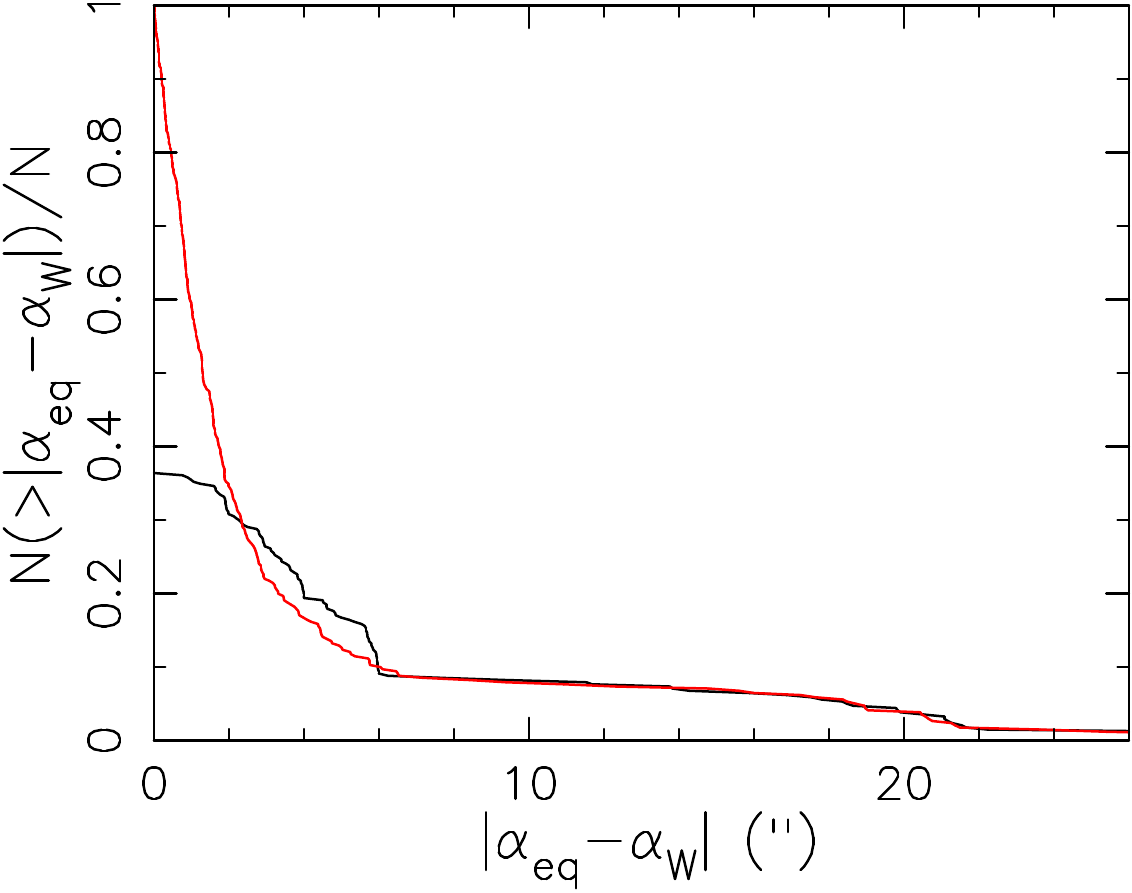}}

\caption{Complement of the normalized empirical cumulative distribution function ECDF,
i.e.\ 1$-$ECDF, for the absolute values of the differences  between
the right ascension $\alpha_\mathrm{eq}$ computed with Eq.\,\ref{e:alpha} and
the catalogued right ascension $\alpha_\mathrm{W}$. Red: vertical axis 
for the exact values $\alpha_\mathrm{eq}$; black: for
the values $\alpha_\mathrm{eq}$ rounded to the nearest allowed integer of $S_\alpha$. 
\label{f:alpha}}
\end{figure}

To check the accuracy of the computation of the right ascension
$\alpha_\mathrm{eq}$ from the right-hand side of Eq.\,\ref{e:alpha}
with the catalogue value $\alpha_\mathrm{W}$, we must take into
account that only 14 values for the number of seconds $S_\alpha$ are
allowed as tabulated values (as explained with Eq.\,\ref{e:gms}).
Thus, each computed value $S_{\alpha_\mathrm{eq}}$ is bracketed by two allowed
values. For the 273 entries with $\phi$ given to a single reference star,
the catalogued value $S_\alpha$ corresponds to the nearest bracketing 
value $S_{\alpha_\mathrm{eq}}$ in 181 cases,
to the other bracketing value in 54 cases, and to neither in 38 cases.

For the 137 entries with $\phi$ given to two or -- in the case of W\,1 --
three reference stars, the catalogued value of $\alpha$ corresponds to the nearest 
bracketing value in 74 cases, to the other bracketing value 
in 28 cases, and to neither in 35 cases. 
In those cases where both values of $\alpha$ are the same after rounding, 
no choice is necessary. If the two (or three) computed values for $\alpha$
are different, the catalogue sometimes gives the value derived from the
average (for example for W\,29); sometimes it gives the value derived
from just one computed value of $\alpha$. We have not been able
to determine a criterion on the basis of which this one value is
chosen: it can be the one found in the lowest or in the highest iteration,
the one found with the largest or smallest angle $\phi$ to the reference 
star, or the one found with the reference stars with the lowest or highest 
absolute value of the declination $|\delta_r|$.

To see whether in computing the right ascensions the
tabulated values of $\delta$ or the values computed from altitudes were used, we
compare the results for $\alpha_\mathrm{eq}$ for the reference star  W\,25 
for both options. This indicates that the tabulated values  were used. 

Figure\,\ref{f:alpha} illustrates the difference distribution between
the catalogued value $\alpha_\mathrm{W}$ and the value
$\alpha_\mathrm{eq}$ computed with Eq.\,\ref{e:alpha}.  
The distributions are shown for the exact values $\alpha_\mathrm{eq}$
computed with Eq.\,\ref{e:alpha} and for the values
$\alpha_\mathrm{eq}$ rounded to the nearest allowed integers of
$S_\alpha$. For entries with more than one $\phi$ we use the
first $\phi$ only, to avoid double counting of one entry in the catalogue;
this leaves 341 values.
The difference between the two curves shows the effect of
rounding: the rounding masks small computational errors when the
catalogued value is the bracketing allowed value closest to the
exactly computed one; it increases the error when catalogued value
corresponds to the other bracketing value. As a result, the median
value of $|\alpha_\mathrm{eq}-\alpha_\mathrm{W}|$ is $1\farcs3$ for 
exact $\alpha_\mathrm{eq}$, but after rounding 217 entries have
$|\alpha_\mathrm{eq}-\alpha_\mathrm{W}|=0$.
The average and rms of $\alpha_\mathrm{eq}-\alpha_\mathrm{W}$ are
0\farcs7 and 6\farcs6 for the exact $\alpha_\mathrm{eq}$ values, and 0\farcs6 and 6\farcs7 
for the rounded $\alpha_\mathrm{eq}$ values.

\begin{figure}
\centerline{\includegraphics[angle=0,width=0.8\columnwidth]{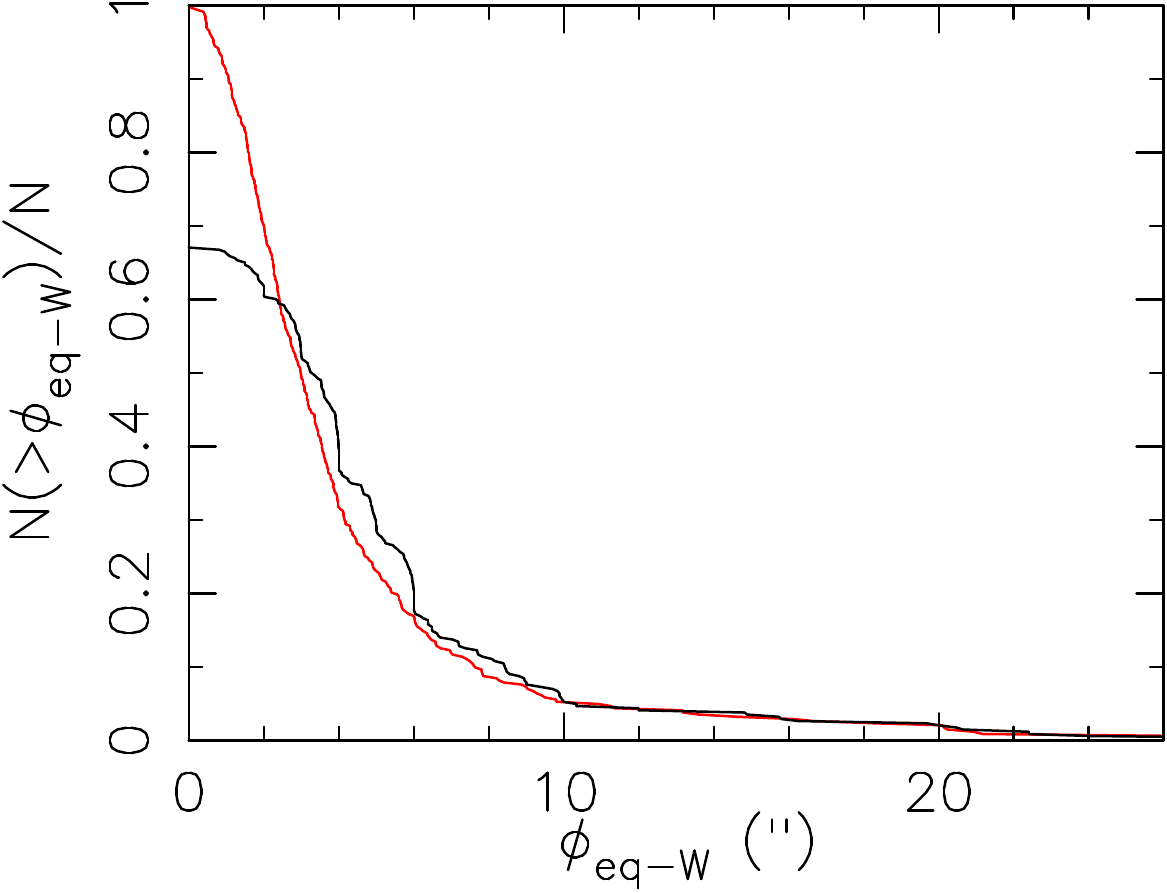}}

\caption{Complement to the normalized empirical cumulative distribution function,
i.e. 1-ECDF, of the differences $\phi_\mathrm{eq-W}$ between
the catalogued position and the position
computed with  Eqs.\,\ref{e:eqtoecl} and \ref{e:eqtoecb}. Red: vertical axis
for the exact computed values $\lambda_\mathrm{eq},\beta_\mathrm{eq}$; black: for
the values rounded to the nearest allowed integer of $S_\lambda,S_\beta$. 
\label{f:compeqb}}
\end{figure}

As we will see below, these errors are negligible with respect to the
measurement errors.

\subsection{Conversion of equatorial to ecliptic coordinates\label{s:eqtoecl}}

The equations which convert equatorial into ecliptic coordinates are
\begin{equation}
\tan\lambda = \frac{\sin\alpha\cos\epsilon+\tan\delta\sin\epsilon}
{\cos\alpha}
\label{e:eqtoecl}\end{equation}
and 
\begin{equation}
\sin\beta=\sin\delta\cos\epsilon-\cos\delta\sin\alpha\sin\epsilon
\label{e:eqtoecb}\end{equation}
We compare the values for $\lambda$ and $\beta$ obtained with these
equations from the equatorial coordinates in column\,5 of \manuscript\
with the values listed in column\,6. In accordance with our finding in 
Sect.\,\ref{s:desc} and Table\,\ref{t:aldeproc}, we set the obliquity
to $\epsilon=23\degr31\arcmin$.

The angular difference $\phi_\mathrm{eq-W}$ between the computed and
catalogued positions in ecliptic coordinates is found with
$$\phi_\mathrm{eq-W} = \phantom{catalogued positions in ecliptic coordinates is found with}$$
\begin{equation}
\phantom{m}2\arcsin\sqrt{\sin^2\frac{\beta_\mathrm{eq}-\beta_\mathrm{W}}{2}+
\cos\beta_\mathrm{eq}\cos\beta_\mathrm{W}\sin^2\frac{\lambda_\mathrm{eq}-\lambda_\mathrm{W}}{2}}
\label{e:havercomp}\end{equation}
Fig.\,\ref{f:compeqb} shows
the complement to the normalized empirical cumulative distribution function ECDF,
i.e.\ 1$-$ECDF, of the differences $\phi_\mathrm{eq-W}$ found with 
Eqs.\,\ref{e:eqtoecl} --\ref{e:havercomp} for the exact computed values of
$\lambda_\mathrm{eq}$ and $\beta_\mathrm{eq}$ and for their values rounded
to the nearest allowed values.
The average and rms values in our first calculation were very high, because
of three outliers W\,55,  W\,214, and W\,374. Comparison of the calculated 
and catalogue positions indicate errors in \manuscript\ for these entries,
and we emend the ecliptic coordinates accordingly (see Sect.\,\ref{s:eclip}).
For 113 entries the rounding leads to $\phi_\mathrm{eq-W}=0$.
The median value of $\phi_\mathrm{eq-W}$ is $3\farcs0$ for 
exact calculation, but after rounding to allowed values it becomes  $3\farcs3$.
The average and rms values are 4\farcs1  and 4\farcs2
for the exact values, and 3\farcs9 and 4\farcs7 for the rounded values.

If we adopt a value of the obliquity 1\arcmin\ higher or lower
than 23\degr31\arcmin\ the average and rms values  of $\phi_\mathrm{eq-W}$
increase strongly, which confirms that the value actually used is
indeed 23\degr31\arcmin.

\begin{figure}
\centerline{\includegraphics[angle=0,width=0.8\columnwidth]{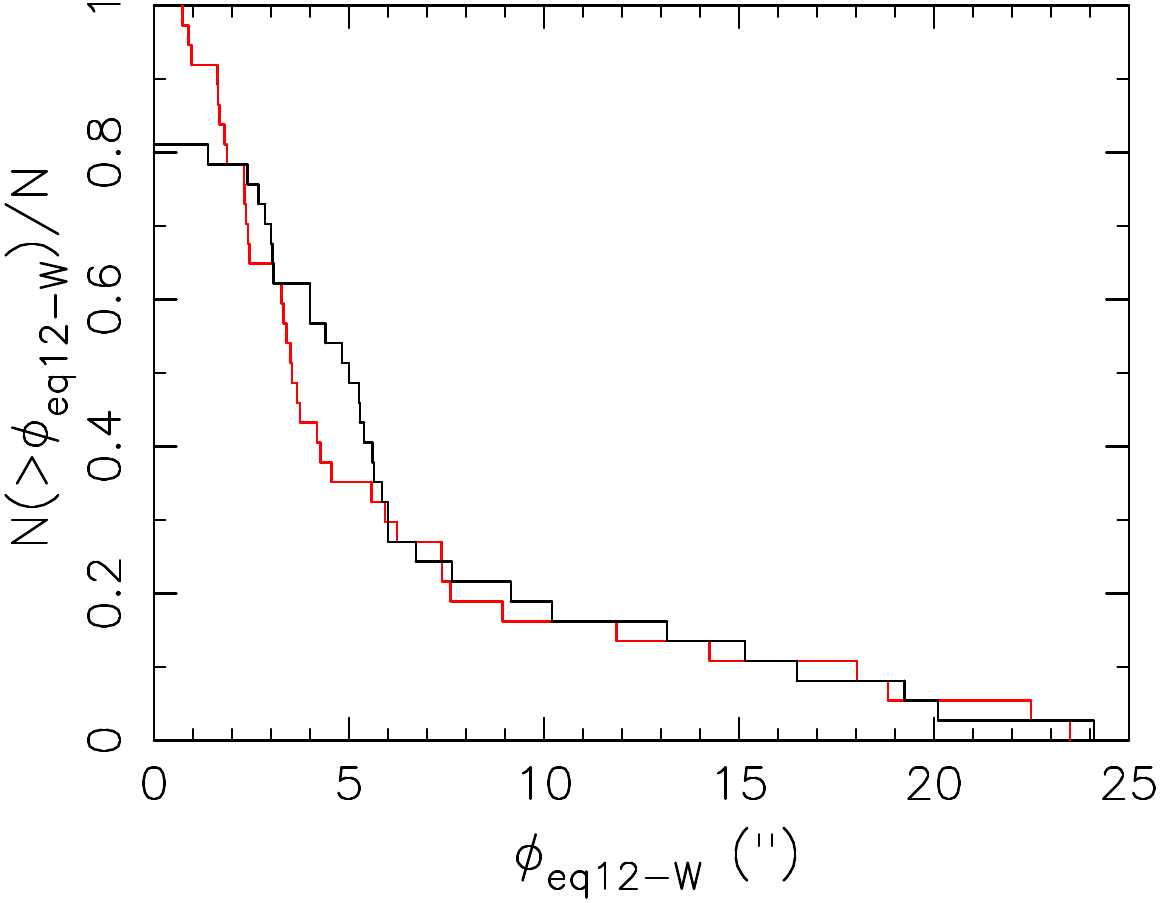}}

\caption{Complement to the 
normalized empirical cumulative distribution function,
i.e. 1-ECDF, of the differences $\phi_\mathrm{eq12-W}$ between
the position computed with  Eq.\,\ref{e:twophi} and the catalogued position. Red: vertical axis
for the exact computed values $\lambda_\mathrm{eq12},\beta_\mathrm{eq12}$; black: for
the values rounded to the nearest allowed integer of $S_\lambda,S_\beta$. 
\label{f:twophi}}
\end{figure}

\subsection{Position from angles with two other stars}

For 41 catalogue entries ecliptic coordinates are given, but no
altitudes or equatorial coordinates. For 37 of these, angles $\phi_i$
are listed to two reference stars with known coordinates
$\lambda_i,\beta_i$, $i=1,2$. Each angle $\phi_i$ defines a circle
around its reference star, and one of the intersections between the
circles gives the previously unknown position $\lambda,\beta$ for
one of the 37 stars.  To calculate these, we rewrite the equivalent of
Eq.\,\ref{e:alpha} in ecliptic coordinates in the form of two
non-linear equations for two unknowns $\lambda$ and $\beta$:
\begin{equation}
F_i(\lambda,\beta) = \cos\beta\cos\beta_i\cos(\lambda-\lambda_i)+\sin\beta\sin\beta_i
-\cos\phi_i = 0
\label{e:twophi}\end{equation}
These can be solved iteratively. Whether this is the method by which the
ecliptic coordinates of the 37 stars were determined is not clear. The
catalogued coordinates of the 4 stars without altitude and with
no or just one angle to another star indicate that not all measured
altitudes and / or angles are catalogued.

If  the coordinates of 37 entries were determined from
Eq.\,\ref{e:twophi}, presumably by trial and error, we can check its
accuracy in two ways. First we compute
$|F_1(\lambda,\beta)|+|F_2(\lambda,\beta)|$ from the values $\lambda$,
$\beta$, $\lambda_i$, $\beta_i$, $\phi_i$ given in the catalogue.
Twenty-two values are less than $10^{-5}$, fifteen lie between
$10^{-4}$ and $10^{-5}$, indicating an excellent accuracy of the
computations.  Second, we can use a modern code to solve
Eq.\,\ref{e:twophi} by numerically minimizing
$|F_1(\lambda,\beta)|+|F_2(\lambda,\beta)|$ as a function of $\lambda$
and $\beta$, and compute the angle $\phi_{\mathrm{eq12-W}}$ between the
optimal computed position $\lambda_{\mathrm{eq12}}$,
$\beta_{\mathrm{eq12}}$ and the catalogued position $\lambda_W$,
$\beta_W$.  As initial guess for the iterative solution of
Eq.\,\ref{e:twophi} we use the catalogued position, and we end the
iteration when $|F_1|+|F_2|<10^{-6}$.  The results are shown in
Figure\,\ref{f:twophi}, $\phi_{\mathrm{eq12-W}}$ for the exactly
computed position, and for the computed position rounded to the
nearest allowed values of $S_\lambda$ and $S_\beta$. The median
is $3\farcs5$ for the exact and $5\farcs0$ for the rounded
solutions.

\section{Accuracy of the measurements}

\begin{figure*}
\centerline{\includegraphics[angle=0,width=0.8\columnwidth]{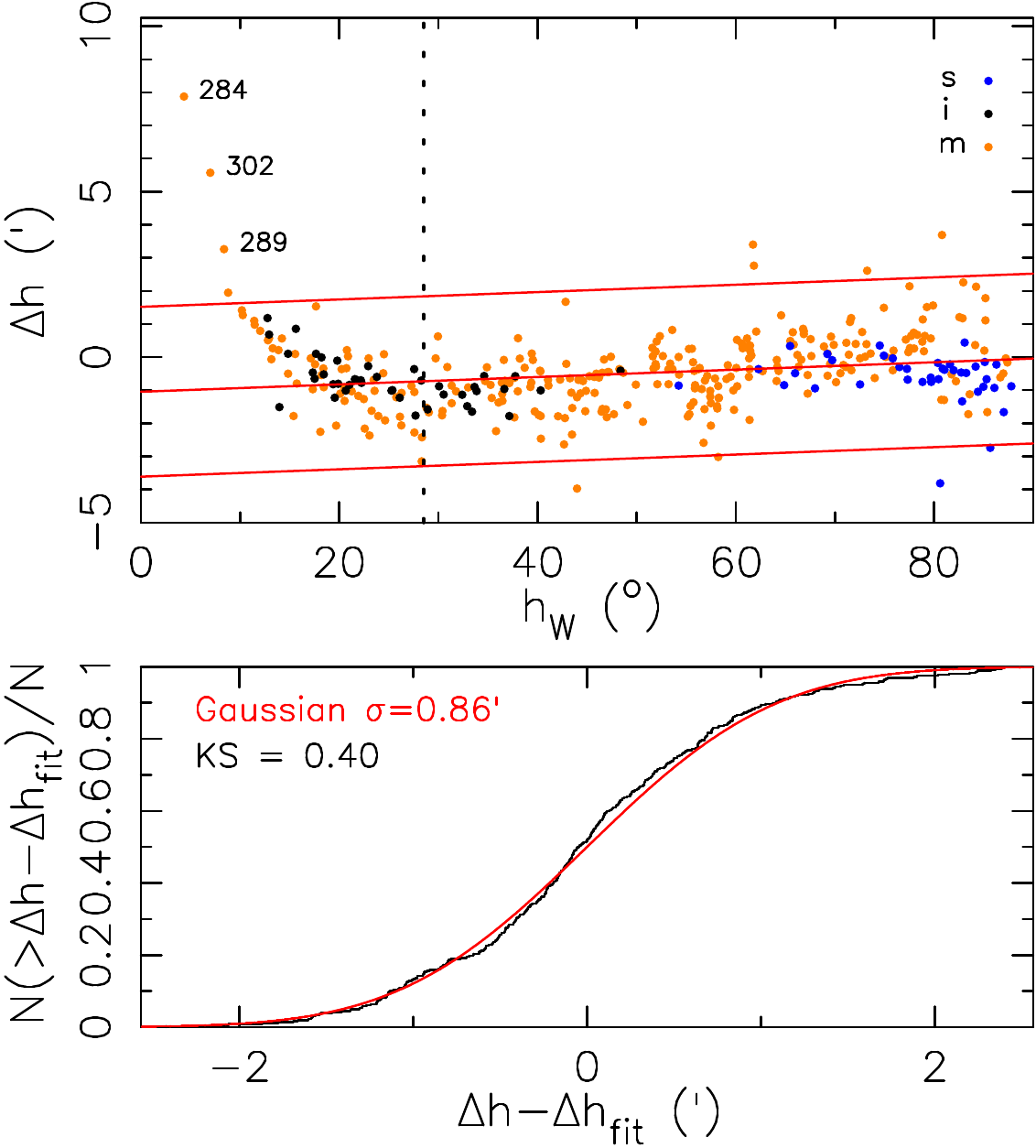}
\hspace*{0.4cm}\includegraphics[angle=0,width=0.8\columnwidth]{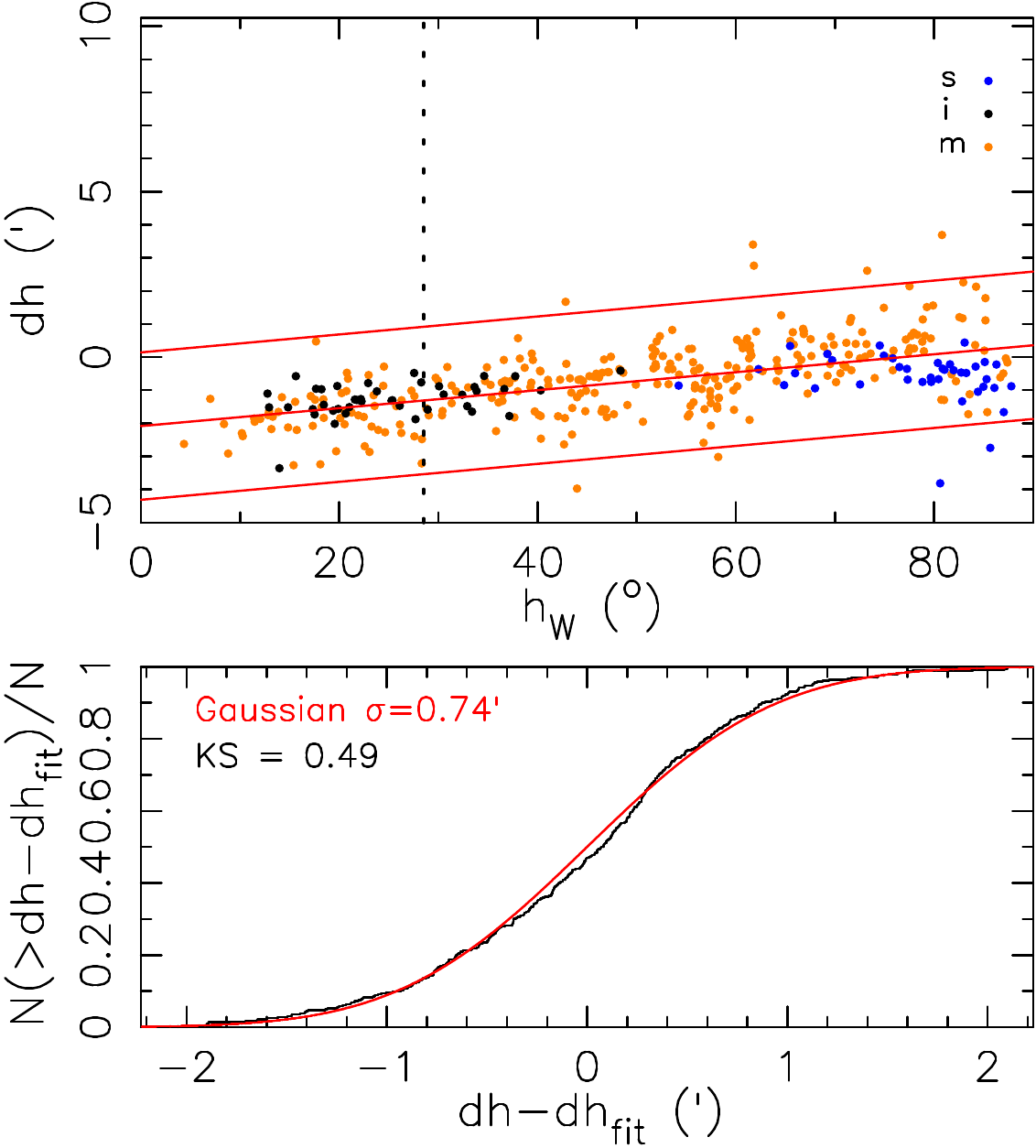}}

\caption{Top: the differences $\Delta h$ (left) and $dh$ (right) between
  the altitudes measured in Kassel and the altitudes that should
  have been measured according to modern data
  (Eqs.\,\ref{e:deltah},\ref{e:deltahb}).  The solid lines show the
  best linear fit and its 3-rms range. Points outside this range
  were ignored in the fit.  $R_\mathrm{R}=0$ and thus $\Delta h = dh$ to the right of the
  dashed vertical line.  Below: cumulative distribution of the
  difference between the measured $\Delta h$ and $dh$ and their best
  fit, compared with the distribution expected for a Gaussian with $\sigma$
  equal to the rms of the differences with the best fit. KS indicates the $p$-value 
  from a two-sided Kolmogorov-Smirnov test that the observed points are drawn from
  the Gaussian \label{f:deltah}}
\end{figure*}

By comparing the catalogued altitudes at culmination and the angles
between stars with values computed from modern HIPPARCOS data and
modern refraction values we can determine the accuracy of  the measurements.

\subsection{Measured altitudes\label{s:measalt}}

\begin{table}
\caption{Results of straight line fits $a+bh$ to $\Delta h$ (Eq.\,\ref{e:deltah}) and
$dh$ (Eq.\,\ref{e:deltahb}) as a function of $h$. The columns indicate which
data points (meridional, superior and inferior culmination) are fitted,  which range of $h$
was included, the intercept $a$ and slope $b$ of the fit, the rms of the differences
between data point and fit, the number of points $n$ used  in the fit, 
and $n_\mathrm{rej}$ the number of points rejected because they lie more than
3 rms from the fitted line.\label{t:deltah}}
\begin{tabular}{|cccrrrrr|}
\hline\hline
fit to:&  & range & $a(\arcmin)$ & $b(\arcmin / \degr)$ & rms & $n$ & $n_\mathrm{rej}$\\ 
\hline
$\Delta h$      & msi &all $h$         & $-$1.0(1) & 0.011(2) &  0\farcm86& 359& 15\\
$\Delta h$      & m   &all $h$         & $-$1.3(1) & 0.016(3) &  0\farcm93& 282& 10\\
$\Delta h$      & si  &all $h$         & $-$0.8(1) & 0.003(2) &  0\farcm54&  78&  4\\
$\Delta h,dh$   & msi &$h\geq29^\circ$ & $-$1.8(2) & 0.023(3) &  0\farcm76& 275& 11\\
$\Delta h,dh$   & m   &$h\geq29^\circ$ & $-$2.1(2) & 0.030(3) &  0\farcm78& 225&  8\\
$\Delta h,dh$   & si  &$h\geq29^\circ$ & $-$1.3(2) & 0.011(3) &  0\farcm46&  50&  3\\
$dh$            & msi &all $h$         & $-$2.1(1) & 0.027(2) &  0\farcm74& 362& 12\\
$dh$            & m   &all $h$         & $-$2.4(1) & 0.034(2) &  0\farcm75& 282& 10\\
$dh$            & si  &all $h$         & $-$1.5(1) & 0.013(2) &  0\farcm43&  78&  4\\
\hline
\end{tabular}
\end{table}

As we have seen in Sect.\,\ref{s:altidecs} the derivation of the
declinations of the entries in \manuscript\ indicates that the
catalogued altitudes $h_\mathrm{W}$ are the true altitudes $h$. The altitude
$h_\mathrm{a}$ measured in Kassel is found from this by adding the
refraction $R_\mathrm{R}$ according to Rothmann (see
Eq.\,\ref{e:altsapp}).  From the modern HIPPARCOS data we determine
the declination in 1586, and from this the actual true altitude
$h_\mathrm{HIP}$ with Eqs.\,\ref{e:geophis}-\ref{e:geophim} to which
the modern value $R$ for refraction is added. Repeated entries are ignored,
as are the two entries representing a cluster: W\,112 and W\,247. Atmospheric refraction,
especially at low altitudes, depends on meteorological conditions, and
in the absence of detailed information can be given only for average
conditions.  For this, we use an equation by S\ae mundsson (1986,
cited in Meeus 1998): \nocite{saemundsson86}\nocite{meeus98}
\begin{equation}
R(\arcmin) = 1.02/\tan\left(h(^\circ)+\frac{10\fdg3}{h(\degr)/1\degr+5.11}\right)
\label{e:refrac}\end{equation}
(The division by 1$^\circ$ in the denominator is explicitly shown to indicate that  the
denominator is dimensionless.)
The difference between the altitudes observed (in Kassel) and predicted (from
HIPPARCOS data) is
\begin{equation}
\Delta h = h_\mathrm{W}+R_\mathrm{R} - (h_\mathrm{HIP}+R)
\label{e:deltah}\end{equation}
In the computation of $h_\mathrm{HIP}$ from the declination we use the modern value 
$\phi_\mathrm{G}=51^\circ18'49\farcs212$; this is 10\farcs788 less than the value 
determined in Kassel -- or about 330\,m on the ground.
If the altitudes tabulated in \manuscript\ are the {\em apparent} altitudes $h_a$
(the rationale for this possibility is given below), comparison with modern data gives
\begin{equation}
dh = h_\mathrm{W}- (h_\mathrm{HIP}+R) = \Delta h -R_\mathrm{R}
\label{e:deltahb}\end{equation}
For $h>29\degr$, where $R_\mathrm{R}=0$,  $dh=\Delta h$.
In Fig.\,\ref{f:deltah} we show $\Delta h$ and $dh$ as a function of
altitude, together with linear fits.  To eliminate undue influence of
outliers, we iteratively determine the rms of the distances to the
best fit, and remove points at distances more than 3 rms. For
three entries the catalogue value is clearly wrong: W\,108, W\,283 and
W\,367 with $dh= 17\farcm4$, $-$49\farcm1 and $-$61\farcm8,
respectively. 
The rms values show that rounding of $S_h$ to allowed values has a
negligible effect, and may be ignored in the analysis.

Figure\,\ref{f:deltah} shows two remarkable features.  First, the
points from superior culmination on average give a lower difference
$\Delta h$ than the points from meridional culminations at the same altitude. 
In addition to results for fits to all points, we therefore show in Table\,\ref{t:deltah} 
the results for fits to data from southern and northern culminations separately.
For fits to all $h$ and for fits to $h>29^\circ$ the dependence of $\Delta h$ and $dh$
on altitude $h$ is significantly shallower for the northern culminations than
for the southern culminations.

The second striking feature of Fig.\,\ref{f:deltah} is the non-zero
slope of the fit to $\Delta h$, that ill describes the points at
$h<29\degr$, which curve upwards towards lower $h$. The results for
the fits of $\Delta h$ to points at $h>29^\circ$ are incompatible
with the fits to all $h$, when data combined from northern and
southern culminations are considered, and also when data from northern
or southern culminations are considered separately.
In contrast the fits of $dh$ to data at $h>29^\circ$ are compatible with
the corresponding fits for all $h$, especially when southern and northern
culminations are fitted separately. A striking illustration of this is
provided by the three points with lowest altitude,  which are well outside
the 3-rms range in the fit for $\Delta h$, but well within the 3-rms range
in the fits for $dh$ (Fig.\,\ref{f:deltah}).

If  only points more than 4 rms are iteratively removed in the fitting procedure,
all values for the intercept $a$ and slope $b$ of the best fits are the same
within the error as those listed in Table\,\ref{t:deltah}. Hence our
conclusions are independent of the exact criterion for inclusion and
rejection of points in the fit.

We are thus faced with the conundrum that the comparison between
the altitudes and declinations in the catalogue indicates that the 
altitudes are the true altitudes, whereas the comparison of the
altitudes with those predicted from modern (HIPPARCOS) data suggests
that they are the apparent altitudes. $\Delta h$ and $dh$
are close to zero at altitude 90\degr\ for the fits to all points and
to those from southern culminations only. This indicates that
the measurement apparatus was aligned to the zenith rather than to
the horizon, understandably in hilly country. Most points from
northern culminations have $\Delta h<0$ and $dh<0$, which may indicate
a misalignment with the zenith for these measurements.
The dependence of $\Delta h$ and $dh$ on altitude $h$
could be due to a inaccurate scale division. The problem with this
explanation is the (near) absence of a dependence of $\Delta h$ on $h$
for the northern culminations. 

\begin{figure}
\centerline{\includegraphics[angle=0,width=0.8\columnwidth]{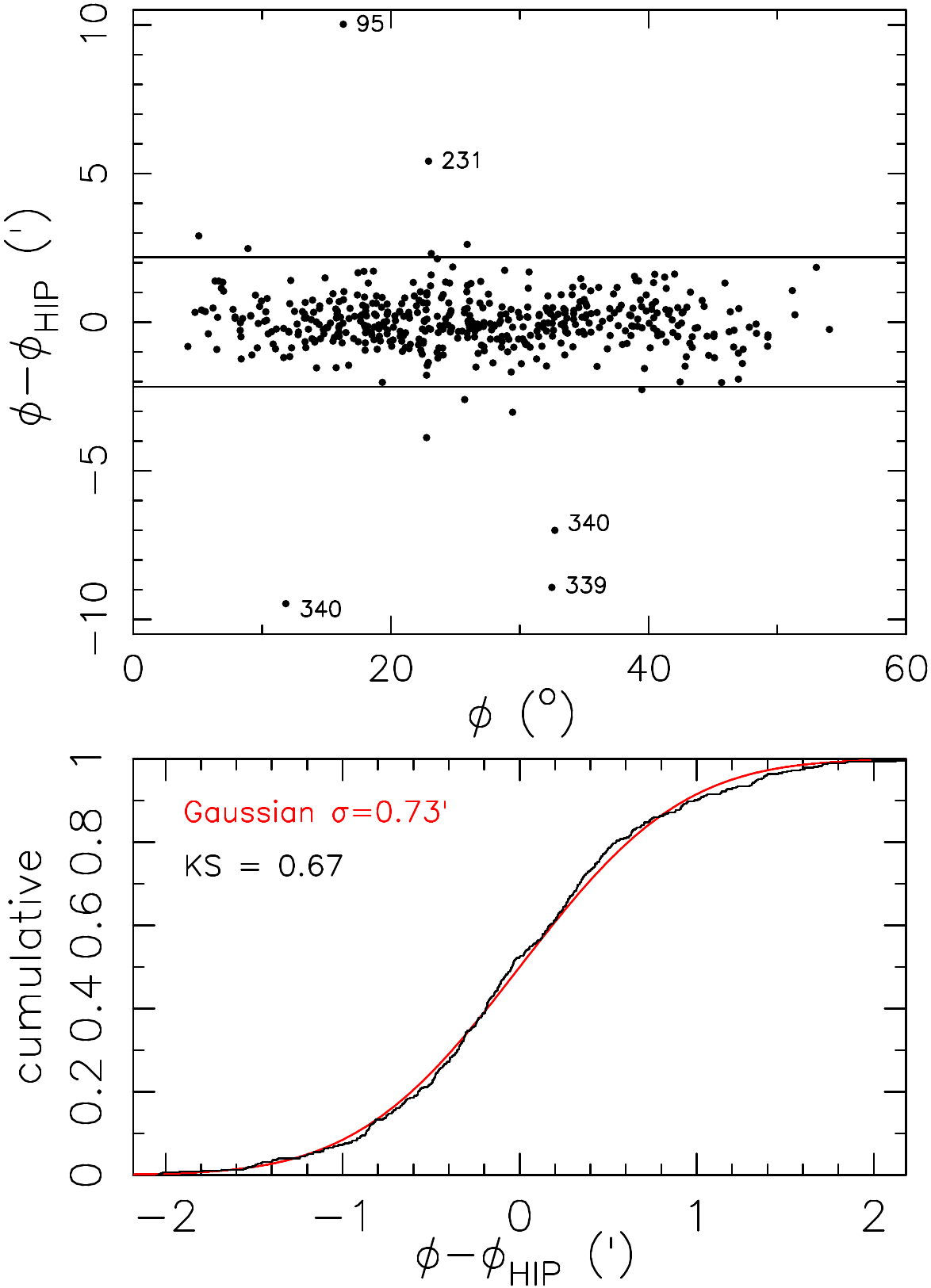}}

\caption{Top: the differences between the angles $\phi$ between two stars as
given in \manuscript\ and the angles $\phi_\mathrm{HIP}$ derived from modern 
(HIPPARCOS) data, as a function of  $\phi$. For some outliers $W$ is indicated
(Table\,\ref{t:dists}). Below: cumulative distribution of
the differences, compared with a Gaussian centered on zero.   \label{f:phi}}
\end{figure}

\subsection{Angles between stars}

\manuscript\ lists the angular distances $\phi$ between stars in column\,2.
From the modern (HIPPARCOS) data we can compute the angle $\phi_\mathrm{HIP}$ at the
time of the observation, and subtract this from $\phi$ to obtain the
measurement error. The results are shown in Fig.\,\ref{f:phi}. There is no
dependence on $\phi$ and the average difference is compatible with
zero. A straight-line fit, iteratively removing outliers at more than 3 rms,
gives 
$$\phi-\phi_\mathrm{HIP}\, (\arcmin)=0.04(9)-0.001(3)\phi (\degr)$$
Assuming that the average value is zero, the rms of $\phi-\phi_\mathrm{HIP}$
is $0\farcm73$, sufficiently large that
rounding of $S_\phi$ to allowed values has a negligible effect.The
furthest outliers are the angles between W\,108-W\,105 and between
W\,339-W\,331, for which $\phi-\phi_\mathrm{HIP}=26\farcm1$ and
$-$20\farcm6, respectively.

\section{Accuracy of the catalogue\label{s:accurcat}}

\begin{figure}
\centerline{\includegraphics[angle=0,width=\columnwidth]{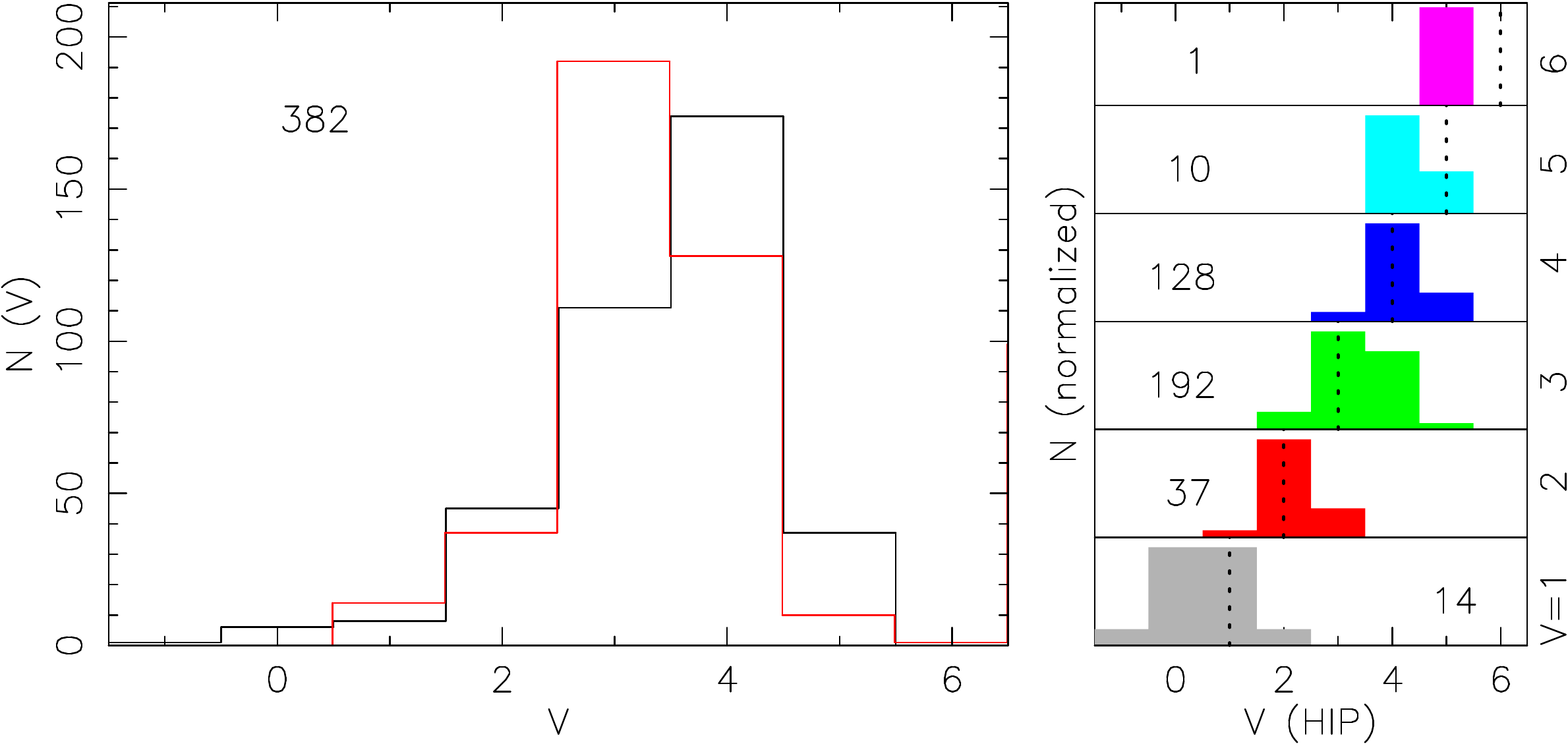}}

\caption{Left: Distribution of the magnitudes given in \basic\ (black) and
of the magnitudes of their counterparts in the HIPPARCOS catalogue (red) 
for all securely identified stars. Right: distribution of the magnitudes of
the HIPPARCOS counterparts,  separately for each magnitude in \basic.
The numbers show the number of stars in each frame.
 \label{f:hesmag}}
\end{figure}

Figure\,\ref{f:hesmag} shows the distributions of the magnitudes of 
the  stars as given in \basic, and of the modern
magnitudes of their counterparts. The magnitudes in the catalogue
correlate well with the modern magnitudes. As expected in a catalogue
of limited size, most entries are bright. For the geographical latitude of 
Kassel, $\phi_\mathrm{G}=51^\circ19'$, stars with $\delta<-38^\circ41'$ are
always below the geometric horizon, the practical observation limit
will be closer to the celestial equator. The southernmost star in
\basic\ is $\epsilon$\,Sgr, at $\delta=-34^\circ20'$,
followed by Fomalhout  at $\delta=-31^\circ41'$.
\basic\ is virtually complete at  (modern) magnitude $V<3.0$ and 
$\delta>-31^\circ$: only $\gamma^2$Sgr, $\gamma$\,Hya, $\gamma$\,Eri,
$\alpha^2$CVn are missing. Extending the magnitude limit to $V<3.5$
shows that 24 more stars with $\delta>-31^\circ$ are missing.

\begin{figure*}
\centerline{\includegraphics[width=7.2cm]{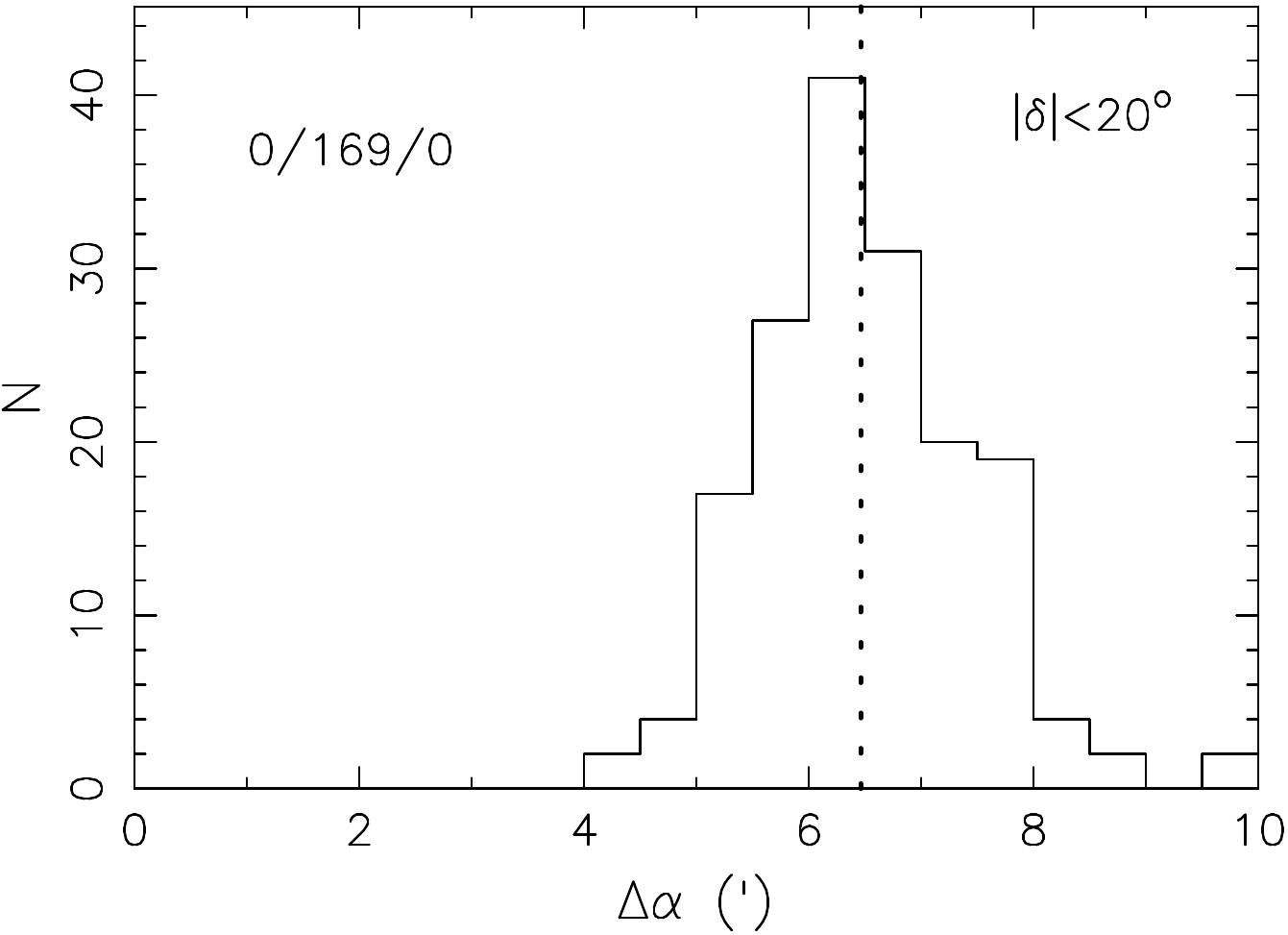}\hspace*{0.6cm}\includegraphics[width=7.cm]{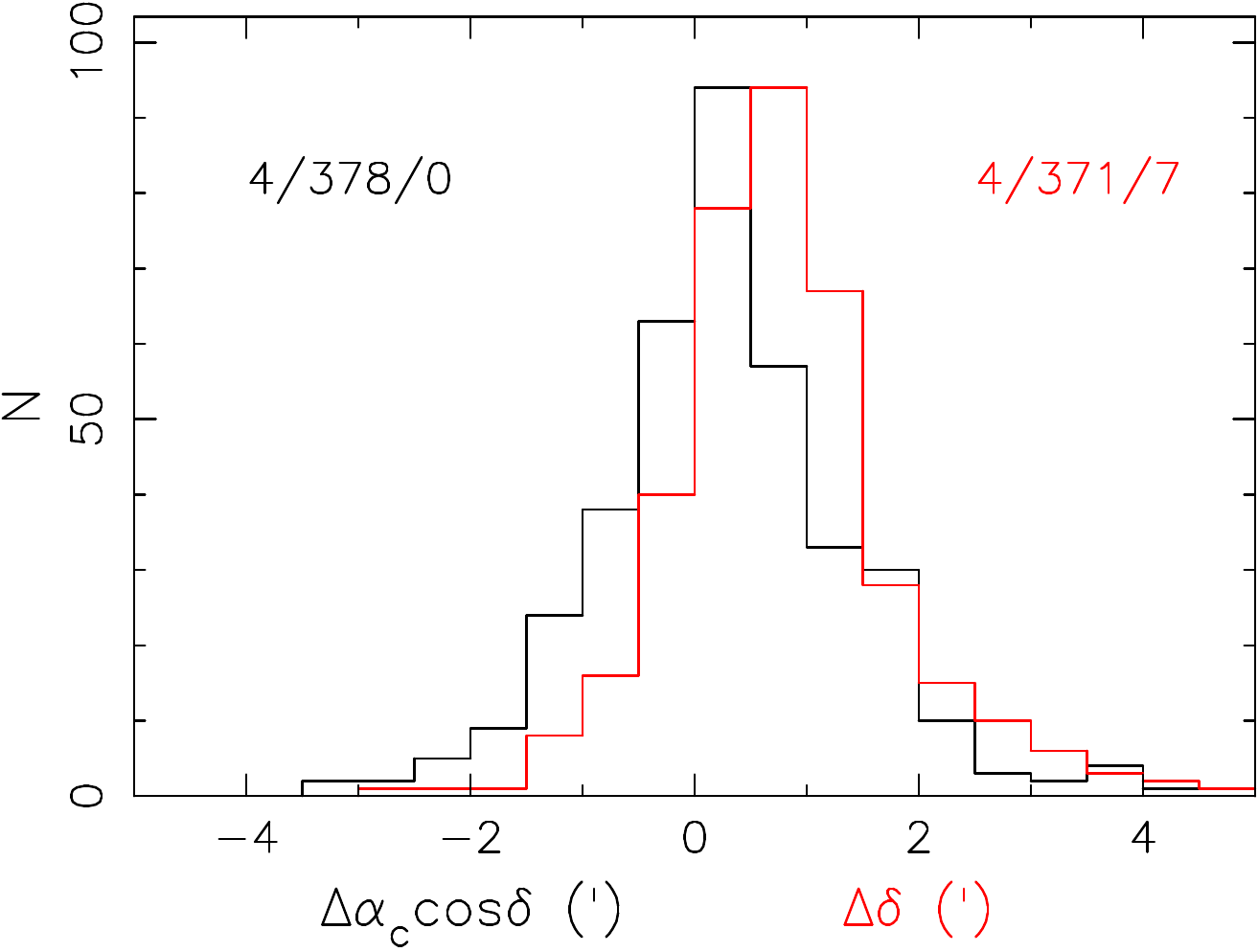}}
\centerline{\includegraphics[width=7.cm]{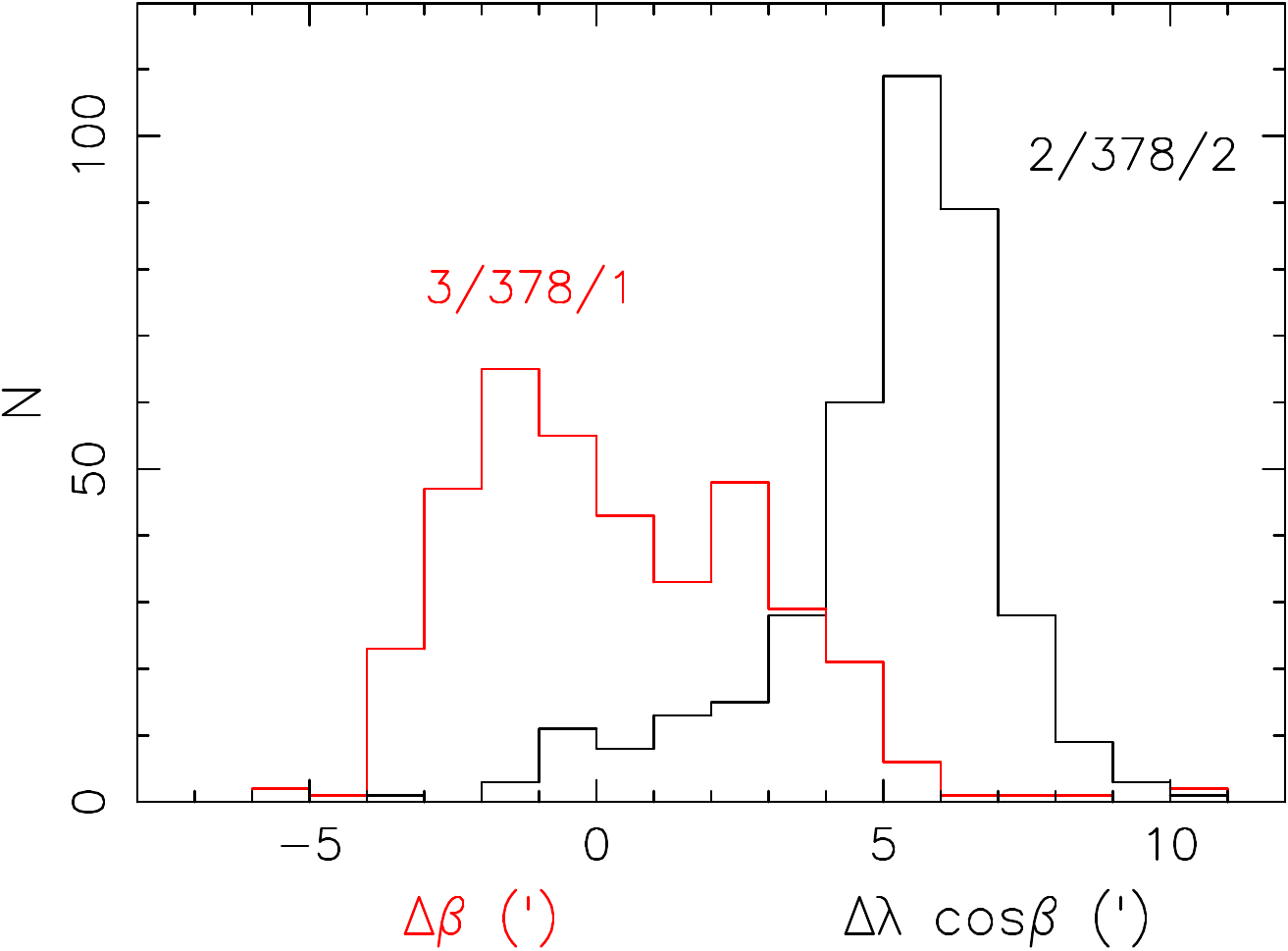}\hspace*{0.6cm}\includegraphics[width=7.cm]{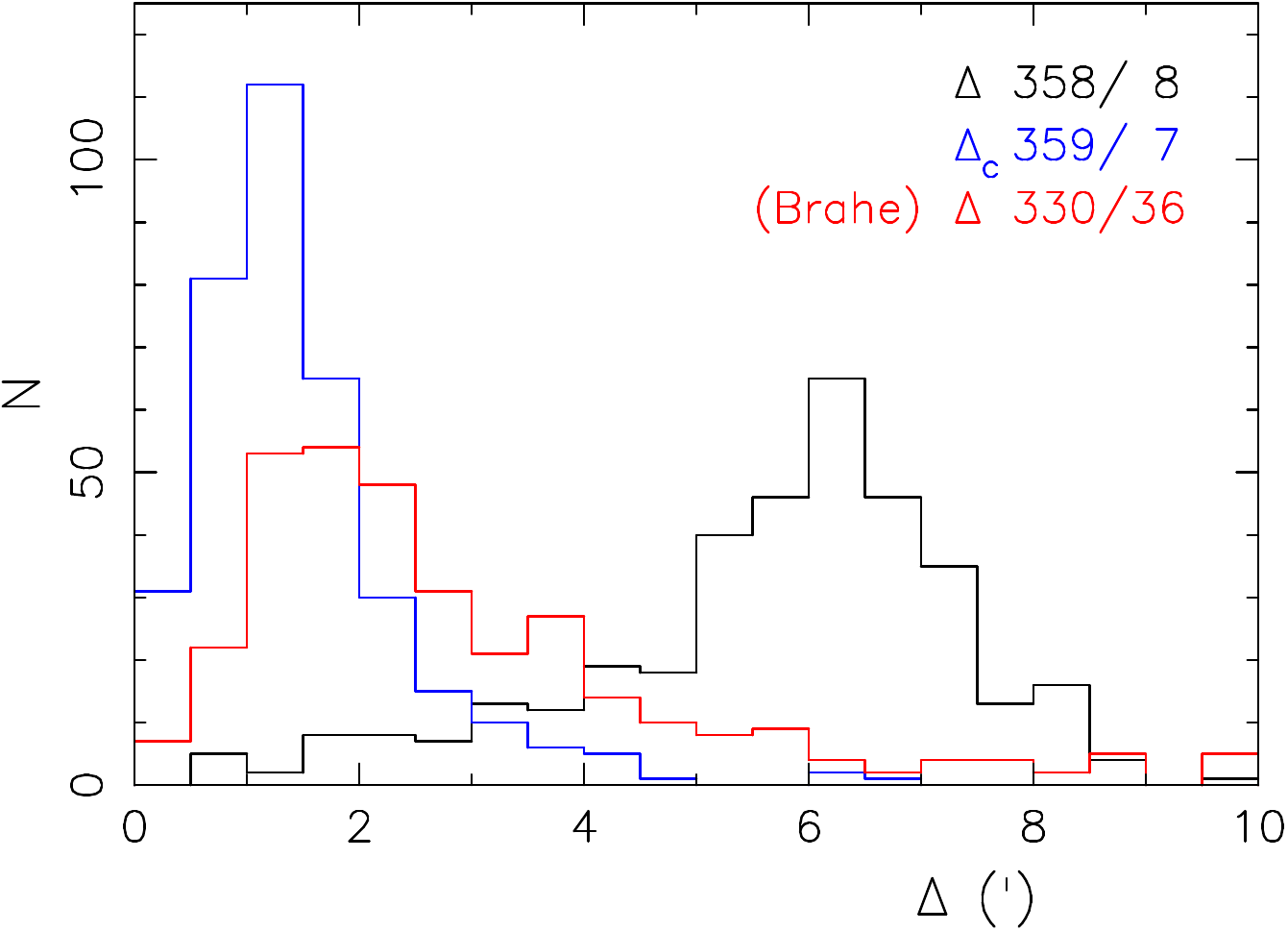}}

\caption{Distributions of various positional errors.
Entries identified with clusters and duplicate entries are not included.
The histograms include entries for which equatorial coordinates are not given in \basic, 
but computed by us from the ecliptic coordinates.
The numbers in the frame indicate the numbers of values 
to the left of / within / to the right of the frame limits.
Top left: Distribution of $\Delta\alpha$ (Eq.\,\ref{e:eqerrors})
for entries with  $|\delta|<20^\circ$.  The average shift, 
indicated with a vertical dashed line, is $6\farcm5$. 
Top right: distributions of $\Delta\alpha_\mathrm{c}\cos\delta$  (black)
and of $\Delta\delta$ (red) --  see Eq.\,\ref{e:eqerrors}.
Below left: Distributions of $\Delta\lambda\cos\beta$ (black) and 
$\Delta\beta$ (red) -- see  Eq.\,\ref{e:eclerrors}.
Below right: Distributions of the position errors $\Delta$ (black, Eq.\,\ref{e:ecldelta})
and $\Delta_\mathrm{c}$  (blue, Eq.\,\ref{e:eqdelta}).
For a fair comparison with the star catalogue of Brahe (edition by Kepler 1627,
position errors shown in red), we select only the 366 entries that are both 
in \basic\ and in Brahe's catalogue.
   \label{f:errors}}
\end{figure*}

The excellent overall accuracy of \basic\ is indicated by the fact
that all entries can be securely identified with HIPPARCOS
counterparts, or in two cases clusters of stars.
In Section\,\ref{s:eqtoecl} we showed that {equatorial coordinates
were accurately converted  into ecliptic coordinates. We therefore
include in our analysis equatorial coordinates which are not in \basic,
computed by us from the ecliptic coordinates.  For small angles
  $\Delta$ between the catalogue and HIPPARCOS positions we have
  approximately $\cos\delta\simeq\cos\delta_\mathrm{HIP}$ and
\begin{equation}
\Delta \simeq  \sqrt{(\Delta\alpha\cos\delta)^2 + (\Delta\delta)^2}
\end{equation}
A measurement error $d$ in the direction of right ascension leads to an
error  $\Delta\alpha\simeq d/\cos\delta$, hence
to a large value for $\Delta\alpha$ at high declinations.
This masks the effect of a systematic offset. For this reason
we show in Figure\,\ref{f:errors} (top left) the distribution of $\Delta\alpha$
(Eq.\,\ref{e:eqerrors}) for angles $|\delta|<20^\circ$ only. 
The average is $\simeq6\farcm5$, which is close to
the offset of $6'$ of the vernal equinox in \basic\
found by Brahe (cf.\ Table\,\ref{t:aldeproc}), and justifies
our correction to  the catalogued right ascensions before
we identify the entry.
Astronomers in the 17th century knew about the offset found by Brahe,
and thus could correct for it. Curtz (1666) approximated the correction
by moving the equinox of the catalogue to seven years later, 1593.
Figure\,\ref{f:errors} (top right) show the distributions of 
$\Delta\alpha_\mathrm{c}\cos\delta$ and $\Delta\delta$, which
reflect the measurement errors. 

\begin{table}
\caption{Average and rms of various distributions for values within frame
limits of Fig.\,\ref{f:errors}.\label{t:errors}}
\centerline{\begin{tabular}{|l|c|c||l|c|c|l|}
\hline 
&  ave  & rms &  &ave & rms & source \\
\hline
$\Delta$ & 5.6 & 1.7 & $\Delta_\mathrm{c}$ & 1.4 & 1.0 & this article \\
& & & $\Delta$ & 2.8 & 2.0 & catalogue Brahe \\
$\Delta\lambda$ & 5.1 & 2.1 & $\Delta\beta$ & 0.3 & 2.6 & this article \\
$\Delta\alpha_\mathrm{c}$ & 0.3 & 1.1 & $\Delta\delta$ & 0.7 & 1.0 & this article\\
$\Delta\alpha_\mathrm{c}$ & 0.2 & 1.2 & $\Delta\delta$ & 0.8 & 1.5  & Rothenberg$^a$\\
\hline
\end{tabular}}
\tablefoot{$^a$in Hamel(2002), for entries with $|\Delta\alpha_\mathrm{c}|<10'$ and
$|\Delta\delta|<10'$. Rothenberg also applies a correction of $6'$ to the right
ascensions; his results for $\Delta\alpha_\mathrm{c}$ and $\Delta\delta$ are very similar 
to ours. Uncharacteristically large errors are due to computing or copying error,
and their contribution to the average and rms is unrealistic. Hence we use only
values within the frame limits in the computation of  average and rms.}
\end{table}

The shift of $6'$ in the right ascensions
leads to an erroneous location of the ecliptic, to the right of the
correct location, in the computation of ecliptic coordinates from
equatorial coordinates (Eqs.\,\ref{e:eqtoecl}, \ref{e:eqtoecb}).
This causes an error of 6$'$ in $\lambda$ at   $\lambda=90^\circ$ or 
$270^\circ$ but to an error smaller by a factor $\cos\epsilon$ at the equinoxes. 
It has no  effect on $\beta$ at $\lambda=90^\circ$ or $270^\circ$, but 
at the vernal and autumnal equinoxes leads to errors of about
$-6'\sin\epsilon\simeq-2\farcm6$ and $+2\farcm6$, respectively.
Thus the  offset in right ascension is reflected in an offset in the
longitudes, visible in the distribution of $\Delta\lambda\cos\beta$, and to
a double peak in   the distribution of $\Delta\beta$
as shown in  Figure\,\ref{f:errors} (left below). The error also increases the
angle $\Delta$ (Eq.\,\ref{e:ecldelta}), which peaks slightly above
$6'$, as shown by the black curve in Figure\,\ref{f:errors} (right
below). The pronounced tails of the distributions of
$\Delta\lambda\cos\beta$ and of $\Delta$ derived from the
(uncorrected) ecliptic coordinates, arise because the effect of the
offset of the vernal equinox decreases with declination with
$\cos\delta$.

\begin{table*}
\caption{Measurement accuracies determined from meridian altitudes 
for the mural quadrant of Brahe, determined by Wesley (1978), and for
\alts, determined in Section\,\ref{s:measalt}, for eight
of the fundamental stars of Brahe. The Table also lists the
errors $d\epsilon$ in the obliquity and $d\phi_\mathrm{G}$ in the
geographical latitude. \label{t:funda}}
\centerline{\begin{tabular}{|c|c|c|c|c|c|c|c|c|c||c|c|}
\hline
& $\alpha$\,Ari& $\alpha$\,Tau & $\mu$\,Gem & $\beta$\,Gem
& $\alpha$\,Leo& $\alpha$\,Vir& $\alpha$\,Aql& $\alpha$\,Peg 
 & $<|dh|>$ &$d\epsilon$ & $d\phi_\mathrm{G}$ \\ \hline
\alts &  $-53''$ & $-32''$ & $-36''$ & $2''$ & $2''$ & $-37''$  & $-28''$ & $-21''$ 
& $26''$& $1'25''$ & $11''$ \\ 
Brahe & $116''$ & $60''$ & $24''$ & $26''$ & $-26''$ & $-22''$  & $-100''$ & $20''$ 
& $49''$ & $1'54''$ & $2''$ \\
\hline
\end{tabular}}
\end{table*}

Correcting the right ascensions by subtracting $6'$ much improves the accuracy of the
catalogue (Table\,\ref{t:errors}), as shown for right ascension and declination in Figure\,\ref{f:errors}
(top right) and for the total error by the blue curve in Figure\,\ref{f:errors}
(below right). To enable a fair comparison with the positional accuracy
of the star catalogue of Brahe -- as edited by Kepler (1627) -- we show the
errors $\Delta$ only for those stars that are present in both catalogues.

\section{Discussion and comparison with Brahe}

The observations of stars made at the castle of Kassel,
under the patronage of Wilhelm\,IV, Landgraf von Hessen-Kassel,
with instruments built by Jost B\"urgi,  reached an
accuracy an order of magnitude better than anything achieved before.
Meridional altitudes and angles between stars were measured with an
accuracy of $44''$. The accuracy of instruments and measurements
profited from the exchanges on these topics between Wilhelm\,IV and Tycho 
Brahe (Hamel 2002). The computations with which the measurements
were converted to celestial coordinates were very accurate as well.
An error of about $6'$ in the position of the vernal equinox, however,
corrupted the celestial coordinates, tabulated for  382 stars and two open 
star clusters in ecliptic coordinates and for 342 stars and one open cluster
in equatorial coordinates. The offset of the vernal equinox was
discovered by Brahe; if correction is made for this offset, the positional
accuracy $\Delta_\mathrm{c}$ of \basic\ is better by a factor two
than that of the star catalogue of Brahe himself (Table\,\ref{t:errors}).

Wesley (1978)\nocite{wesley78} has determined the accuracy of the the
measurements with various instruments by Brahe. He determined the
measurement accuracy for eight fundamental stars, used by Brahe as
reference for measurements on other stars, by comparing the meridian
altitudes given by Brahe with modern computation.  This is what we
did in Section\,\ref{s:measalt} for \alts. Our results are compared
with those by Wesley for the mural quadrant of Brahe in
Table\,\ref{t:funda}.  The measurement errors in \alts\ are on the whole
smaller, averaging about half the errors of the mural quadrant of Brahe.

The star catalogue \basic\ is corrupted by the offset of the
vernal equinox. Thus, the positional errors in \basic\ are larger
than those for most of the stars in Brahe's catalogue.
The catalogue by Brahe, however, is corrupted for a significant
fraction of its entries by errors in converting measurement to
celestial positions, e.g.\ by confusing measurements of
different stars and by scribal errors, as shown by Rawlins 
(1993).\nocite{rawlins93}
Of the 366 stars common to \basic\ and the star catalogue of Brahe 
(Kepler 1627) no less than 36 have a positional error larger
than $10\arcmin$ in Brahe's catalogue, but only 8 in even the uncorrected
version of \basic.
It is therefore a moot question which of the two catalogues
is more accurate.
However, when we consider the positional errors in \basic\
after correction for the offset in right ascensions, the combination
of more accurate measurements and more diligent computing
causes \basic\ to be more accurate (Figure\,\ref{f:errors} below right).

The value of the obliquity used in \basic\ for the conversion of
ecliptic coordinates into equatorial ones, or vice versa, is more
accurate than the value $\epsilon(1601)=23^\circ21'30''$ used by
Brahe.  The geographical latitudes used in \basic\ and by Brahe are
both remarkably accurate, and have no consequence for the conversion
of altitudes into declinations. This conversion is affected, however,
by the erroneous values  used for atmospheric refraction.

Wilhelm\,IV and his collaborators did not publish their work, as we
mentioned in the Introduction, but it is interesting to know that the
accuracy of Brahe was not unique.  Of course, Brahe published not only
his star catalogue, but many other works on the renovation of
astronomy.  The accuracy of his planetary positions enabled Kepler to
show that Mars moves in an ellipse with the Sun in a focal point, and
by extension that all planets do. Thus Brahe and Kepler together
revolutionized astronomy.

To our knowledge, \manuscript\ is the first star catalogue to give
both equatorial coordinates and  ecliptic coordinates. Earlier star
catalogues, by Ptolemaios ($\pm$150), al-Sufi ($\pm$964) and Ulugh
Beg (1437), as well as the later star catalogue of Brahe (1627),
only provided ecliptic coordinates. Riccioli (1651) and
Hevelius (1690) also provide both ecliptic and equatorial coordinates.  De
Houtman (1603) gives the first printed star catalogue with equatorial
coordinates (but no ecliptic coordinates).

\begin{acknowledgements}
  We thank Erik H\o g for instigating this work.  This research has made
  use of the SIMBAD database, operated at CDS, Strasbourg, France.
  Computations by FV use Numerical Recipes by Press et al.\ (1986).
  Computations by AS use Astropy, a community--developed core Python package for
 Astronomy (Astropy Collaboration 2013\nocite{astropy2013}) and 
 The Naval Observatory Vector Astrometry Software (NOVAS) for astrometric quantities 
 and transformations (Bangert 2011\nocite{novas2011}). 
\end{acknowledgements}

\begin{appendix}
\section{The HIPPARCOS catalogue for epoch 1586.\label{s:hipparcos}}

For a number of stars the original HIPPARCOS catalogue did not
find a solution for the proper motion. 
These stars are not present in the revised HIPPARCOS-2 catalogue 
by Van Leeuwen (2007).\nocite{leeuwen07} 
We use the position of these stars as given in the HIPPARCOS
catalogue (ESA 1997)\nocite{esa} and collect their proper
motions from various sources, as listed in Table\,\ref{t:addhip} .

\begin{table}[h!]
\caption{Stars added by us to the HIPPARCOS-2 Catalogue\label{t:addhip}}
\center{
\begin{tabular}{rcl@{\hspace{0.2cm}}rr@{ }rl}
\hline\hline
HIP & $V$ &$\alpha(2000)$ & $\delta$(2000) & $\mu_\alpha$\phantom{0} & $\mu_\delta$\phantom{0} \\
 & & (hr)\phantom{53} & (\degr)\phantom{639}  & \multicolumn{2}{c}{(mas/yr)}\\
\hline
 31067 & 6.2 & \phantom{2}6.519444  &       16.938748  & $-$13.30 &  $-$48.23 & a \\
 52800 & 6.9 & 10.794485 & $-$15.243544  &  16.47 &  $-$43.81  & a \\
 55203 & 3.8 & 11.3031 &  31.5308\phantom{44} & $-$453.70 & $-$591.40 & b \\
 59273 & 6.7 & 12.1581 & $-$11.8542\phantom{44} &   0.70 &  $-$69.30 &c \\
 78727 & 4.2 & 16.0727 & $-$11.3736\phantom{44}  & $-$63.2 &   $-$27.0 & d  \\
 80579 & 6.7 &16.4528 & $-$47.5491\phantom{44} & $-$6.1 & $-$20.4 & e  \\
115125 & 5.2 & 23.3185 & $-$13.4548\phantom{44} & 302.\phantom{4} & $-$92.\phantom{4} & d \\
\hline
\end{tabular}}
\tablefoot{a.\ notes in printed version of HIPPARCOS (ESA 1997)\nocite{esa}; 
b.\ UCAC4 (Zacharias et al.\  2012)\nocite{zacharias12} 
c.\ Tycho-2 (H\o g et al.\  2000)\nocite{hoeg00} 
d.\ PPM (R\"oser \& Bastian 1988, Bastian \&\ R\"oser 1993)\nocite{roeser88}\nocite{bastian93}, 
e.\ GAIA-DR1 (Gaia Collaboration 2016)\nocite{gaia16} 
}
\end{table}

To take into account that some HIPPARCOS stars cannot be separated by
the human eye, we proceed as follows. We select all stars in the original
HIPPARCOS catalogue with $V<7.0$, and read the positions of these stars
in the revised HIPPARCOS-2 Catalogue. We  correct these positions for
proper motion over the interval between the HIPPARCOS epoch 1991.25
and the \manuscript\ epoch 1586 and for precession from HIPPARCOS
equinox 2000.0 to \manuscript\ equinox 1586, with Eqs\,3.211 and
3.212-2 of Seidelman (1992).  For each star we find all stars within
$2'$ at that epoch, and merge them into one entry. Thus 80 pairs and
two triples are merged into 82 entries. To determine the position of
the merged entry, the positions of its components are weighted with
their fluxes, separately for the right ascension and for the
declination. The magnitude of the merged entry is found by adding the
fluxes of its components. Thus
with $f_i$ the flux normalized to the flux of a star with $V=0$:
\begin{equation}
f_i= 10^{-0.4V_i};\, \alpha= \sum_i (f_i\alpha_i) / \sum_i f_i;\,
\delta = \sum_i (f_i\delta_i)/\sum_i f_i
\end{equation}
and
\begin{equation}
V = -2.5 \log \sum_i f_i
\end{equation}

\section{Emendations to \manuscript}

\subsection{Angular distances between stars\label{s:emangles}}

\noindent W\,36: the name of the reference star is not given.
The angular distance indicates W\,24, which we enter in
\dists.

\noindent Following W\,165, In Cauda (in the tail) sc.\ of Aquila, is
an entry Extrema Caudae (the end of the tail), which is indicated in
column 7 as {\em in tabulis non extat} (not present in the tables sc.\ of
Ptolemaios).  Two angles to reference stars are given for this entry, towards Aquila
(W\,163) and towards Rostrum Cygni (the beak of the Swan), which we
identify with W\,89. These angles imply that Extrema Caudae
corresponds to HIP\,93244 ($\epsilon$\,Aql). That star indeed is not in the
star catalogue of Ptolemaios.

\noindent W\,340, W\,341: Canis maior (W\,353) is indicated
as the reference star, but is excluded by the angular distance; 
we emend to Canis minor (W\,352)

\subsection{Altitudes\label{s:emaltids}}

\noindent W\,64: we emend 18\degr14\arcmin to
58\degr14\arcmin 

\subsection{Ecliptic coordinates\label{s:eclip}}

On the basis of Sect.\,\ref{s:eqtoecl} we emend

\noindent W\,55: $\beta$ 55\degr41\arcmin10\arcsec\ to
54\degr41\arcmin10\arcsec.

\noindent W\,214: $\beta$ 02\degr49\arcmin54\arcsec\ to
 02\degr47\arcmin54\arcsec.

\noindent W\,374: $\lambda$ \virgo\,2$^\mathrm{d}$42$^\mathrm{m}$50$^\mathrm{s}$ to
 \virgo\,2$^\mathrm{d}$41$^\mathrm{m}$50$^\mathrm{s}$

\section{Notes on individual identifications}

Merged with is abbreviated with m.w.

\noindent W\,26: HIP\,85829 ($V=4.84$): m.w. HIP\,85819 ($V=4.89$)
 
\noindent W\,32: HIP\,86614 ($V=4.57$): m.w. HIP\,86620 ($V=5.81$)

\noindent W\,53: HIP\,110991 ($V=4.07$): m.w. HIP\,110988 ($V=6.31$)

\noindent W\,59: HIP\,75411 ($V=4.31$): m.w. HIP\,75415 ($V=6.51$)

\noindent W\,73: HIP\,79043 ($V=5.00$): m.w. HIP\,79045 ($V=6.25$)

\noindent W\,83: HIP\,91971 ($V=4.34$): m.w. HIP\,91973 ($V=5.73$)

\noindent W\,89: HIP\,95947 ($V=3.05$): m.w. HIP\,95951 ($V=5.12$)

\noindent W\,99: we choose HIP\,99675 (31\,Cyg, $V=3.8$, $\Delta=3\farcm4$) as the counterpart
above the closer, but fainter HIP\,99639 (30\,Cyg, $V=4.8$, $\Delta=2\farcm4$) 

\noindent W\,108: we choose HIP\,5542 ($\theta$\,Cas, $V=4.3$, $\Delta=31\farcm7$)) as the counterpart
above the closer, but fainter HIP\,5536  ($\rho$\,Cas, $V=5.2$, $\Delta=24\farcm0$). 

\noindent W\,156: HIP\,92946 ($V=4.62$): m.w. HIP\,92951 ($V=4.98$)

\noindent W\,174: HIP 102532 ($V=4.27$): m.w. HIP\,102531 ($V=5.15$)

\noindent W\,251: HIP\,43103 ($V=4.03$): m.w. HIP\,43100 ($V=6.58$)

\noindent W\,277: HIP\,78820 ($V=2.56$): m.w. HIP\,78821 ($V=4.90$)

\noindent W\,283: notwithstanding the large positional error $\delta=42\farcm8$ we consider
the identification secure.

\noindent W\,312: HIP\,5737 ($V=5.21$) m.w. HIP\,5743 ($V=6.44$)

\noindent W\,317: HIP\,5131 ($V=5.33$): m.w. HIP\,5132 ($V=5.55$)

\noindent W\,340: HIP\,26220 ($V=4.98$): m.w. HIP\,26221 ($V=5.13$) and
HIP\,26224 ($V=6.71$. The original HIPPARCOS
catalogue (ESA 1997)\nocite{esa}  lists $V$=4.98 for HIP\,26220, but
error 0.00  for $B$$-$$V$; this may indicate that the photometry is
suspect. Ducati (2002)\nocite{ducati02} gives  $V=6.73$.
Since HIP\,26220 is a Herbig Ae/Be star, and thus variable, its
magnitude is probably variable. 

\noindent W\,367: the positional error $\Delta=60\farcm7$ strongly suggests an
error of $1^\circ$ in writing or copying the correct result.

\end{appendix}

\end{document}